\documentclass[twocolumn]{aastex701}

\usepackage{multirow}

\begin{document}

\title{Revisiting the \textit{XMM-Newton} Observations of the Galactic Microquasar SS~433: Implications for the Origin of the Ultrahigh-Energy Emission Detected by LHAASO}

\author[orcid=0009-0004-0470-7457]
{Chao-Nan Tong}
\affiliation{School of Astronomy and Space Science, Nanjing University, 163 Xianlin Avenue,\\
Nanjing 210023, People's Republic of China}
\affiliation{Key Laboratory of Modern Astronomy and Astrophysics,\\
Nanjing University, Ministry of Education, Nanjing 210023, People's Republic of China}
\email{cn_tong@smail.nju.edu.cn}

\author[orcid=0009-0004-7502-7037]
{Hong-Bin Tan}
\affiliation{School of Astronomy and Space Science, Nanjing University, 163 Xianlin Avenue,\\
Nanjing 210023, People's Republic of China}
\affiliation{Key Laboratory of Modern Astronomy and Astrophysics,\\
Nanjing University, Ministry of Education, Nanjing 210023, People's Republic of China}
\email{hbtan@smail.nju.edu.cn}

\author[orcid=0000-0003-1576-0961]
{Ruo-Yu Liu}
\affiliation{School of Astronomy and Space Science, Nanjing University, 163 Xianlin Avenue,\\
Nanjing 210023, People's Republic of China}
\affiliation{Key Laboratory of Modern Astronomy and Astrophysics,\\
Nanjing University, Ministry of Education, Nanjing 210023, People's Republic of China}
\affiliation{Tianfu Cosmic Ray Research Center, Chengdu 610000, Sichuan, People’s Republic of China}
\email{ryliu@nju.edu.cn}

\correspondingauthor{Ruo-Yu Liu}
\email{ryliu@nju.edu.cn}








\begin{abstract}

Recently, the Large High Altitude Air Shower Observatory (LHAASO) detected ultrahigh-energy (UHE; photon energy $E>100$\,TeV) $\gamma$-ray emission toward SS~433, the microquasar embedded in the W50 nebula, making it a promising Galactic PeVatron candidate. We reanalyze the archival \textit{XMM-Newton} observations covering the bipolar jets and the thermal X-ray shell north of SS~433, and derive spatially resolved profiles of the nonthermal X-ray intensity and photon index along both jets. The jet emission softens with distance from the source, implying a correspondingly evolving electron population. In particular, a hard electron component appears close to the jet bases, which can account for the UHE emission from SS~433 via inverse Compton radiation if the magnetic field remains approximately uniform along the jets. The result, however, is highly sensitive to the magnetic field profile. For flux-conserving configurations in which the field decreases as the jet expands, the stronger field required in the inner regions may reduce the number of X-ray-emitting electrons and suppress their inverse Compton emission. Furthermore, electron transport calculations show that injection only at the jet bases cannot reproduce the observed intensity and spectral evolution, particularly the downstream re-brightening features, indicating additional particle injection and/or re-acceleration within the jets.

\end{abstract}



\section{Introduction}

Galactic PeVatrons, sources capable of accelerating particles to PeV ($1\,\rm PeV=10^{15}\,$eV) energies, hold the key to unraveling the origin of Galactic cosmic rays (CRs). Identifying such extreme accelerators is crucial for explaining the distinctive ``knee'' feature observed in the CR spectrum. These sources are primarily probed via their $\gamma$-ray emissions in the ultrahigh-energy (UHE; photon energy $E>100$\,TeV) regime. While supernova remnants (SNRs) have long been regarded as the leading PeVatron candidates, yet showing acceleration up to the PeV range remains challenging. Recent detections of $E>100$\,TeV $\gamma$-rays from several microquasars by the Large High Altitude Air Shower Observatory (LHAASO) and the High-Altitude Water Cherenkov Observatory (HAWC) \citep{2024ApJ...976...30A,2024Natur.634..557A,2024ApJS..271...25C,10.1093/nsr/nwaf496} have therefore highlighted microquasars as a promising class of Galactic PeVatron candidates.

Among them, the microquasar SS~433 stands out as a unique laboratory \citep{doi:10.1142/S0218271810016646,2015ApJ...807L...8B,2018Natur.564E..38A,2019A&A...626A.113S,2020ApJ...889L...5F}. Located at a distance of $d\sim5.5$\,kpc \citep{2004ApJ...616L.159B,2007MNRAS.381..881L}, it is widely believed to host a stellar-mass black hole \citep{2002ApJ...578L..67G,2004ASPRv..12....1F,2018A&A...619L...4B,2019MNRAS.485.2638C,2021MNRAS.507L..19C} and launches a pair of mildly relativistic jets with bulk speed $v\sim0.26c$. In the very-high-energy (VHE) energy band, the High Energy Stereoscopic System (H.E.S.S.) found emission associated with the eastern and western X-ray jets; on both sides, the $\gamma$-ray centroid shifts toward the central binary as the photon energy increases from 0.8\,TeV to above 10\,TeV \citep{2024Sci...383..402H}. This behavior is naturally explained by radiative cooling of accelerated electrons/positrons (hereafter ``electrons''), favoring a leptonic inverse-Compton (IC) origin for the sub-TeV to tens-of-TeV emission. More recently, LHAASO has reported likewise observations in $1-100$\,TeV \citep{10.1093/nsr/nwaf496}, consistent with the H.E.S.S. results; resolving two point-like sources coincident with both jets. At energies $>100$\,TeV, however, LHAASO has detected extended $\gamma$-ray emission spatially correlated with an H\textsc{i} atomic cloud \citep{2020NatAs...4.1177L,2018ApJ...863..103S}. The origin of this UHE component remains unclear. In particular, it is still debated whether the $E>100$\,TeV $\gamma$-rays arise from the IC scattering of ambient photons by electrons freshly accelerated to the highest energies at shocks near the bases of eastern and western jets (possibly at other acceleration sites), or from hadronic $\gamma$-ray production via $pp$ interactions between relativistic protons and the associated atomic cloud. X-ray observations provide a powerful diagnostic to distinguish between these two scenarios. Within the leptonic framework, the same population of high-energy electrons responsible for the $\gamma$-rays via IC scattering will also emit synchrotron X-rays in the magnetic field. Therefore, high-resolution X-ray observations are essential, enabling morphological localization of the acceleration sites and providing constraints on the magnetic field strength.

The X-ray investigation of SS~433 can be traced back to as early as the 1980s–1990s, X-ray missions such as \textit{Einstein} \citep{1983ApJ...273..688W}, \textit{ASCA} \citep{1994PASJ...46L.109Y,2000AdSpR..25..709N} and \textit{ROSAT} \citep{1996A&A...312..306B,1997ApJ...483..868S} carried out large-scale observations of the SS~433 system. These data have already resolved the east and west jets in X-rays, with emissions detected beyond $\sim 15^\prime$ and peaking at $\sim 35^\prime$ from the central binary. The jet spectra are predominantly nonthermal and soften with increasing distance from SS~433. Advances in X-ray instrument have significantly improved angular and spectral resolution, enabling detailed studies of the SS~433 system \citep{2005AdSpR..35.1062M}. \citet{2007A&A...463..611B} analyzed \textit{XMM-Newton} data toward eastern jet of SS~433 and found that the peaking knot consists of two components: a nonthermal power-law component with the photon index of $\Gamma \sim$ 2.17, along with a thermal component characterized by $kT \sim$ $0.2-0.3$\,keV. In addition to the refined spectral and morphological characterization of the eastern jet, \citet{2022ApJ...935..163S} combined \textit{XMM-Newton} and \textit{NuSTAR} data to extend the analysis into the hard X-ray band, showing that nonthermal emission from the jet base, located at $\lesssim 18^\prime$ east of SS\,433, referred to as ``head'', is detected up to 30\,keV. This hard X-ray ``head'' has been identified as a likely particle acceleration site within the eastern jet. To investigate X-ray substructure in the western jet of SS~433, \citet{2022PASJ...74.1143K} analyzed \textit{Chandra} observations of the western region, resolving a particle acceleration site at the jet base and localized re-brightening knots, analogous to the features observed in the east. They derived photon index and intensity profiles for the nonthermal emission along the precession axis and adopted an advection-dominated kinetic model, together with local magnetic field enhancement at knots \citep{2020ApJ...889..146S} to explain the jet’s spatial evolution. In a subsequent study, \citet{2025PASJ...77..880K} conducted a detailed \textit{XMM-Newton} radial profile analysis along the eastern jet and, in combination with the western results \citep{2022PASJ...74.1143K}, applied the same theoretical framework described in \citet{2022PASJ...74.1143K} to both sides, thereby accounting for the east and west evolution within a unified model. However, outside the knot regions, the eastern and western jets displayed different spectral softening trends. If synchrotron cooling were dominant, both sides would be expected to exhibit a comparable, abrupt spectral steepening downstream of the magnetic field enhancement. We attribute this discrepancy to differences in background treatment, and the details are presented in Section~\ref{jet spectrum}. Importantly, recent X-ray studies have not been limited to the two-sided jets. \citet{2024ApJ...975L..28C} reported evidence for thermal X-ray emission in the northern shell region of SS~433 and interpreted it as a SNR, implying a potential particle-acceleration site that may also contribute to the UHE emission.

Motivated by previous studies and open questions regarding the UHE $\gamma$-ray origin, this work utilizes a more comprehensive \textit{XMM-Newton} coverage of SS~433 to provide a multiwavelength perspective on this system. By analyzing the spatial and spectral distribution of X-ray emitting electrons, we aim to quantify the leptonic contribution to the $>100$\,TeV emission detected by LHAASO, and subsequently discuss the possible scenario of hadronic origin. Combining these X-ray results with $\gamma$-ray data allows us to discriminate between leptonic and hadronic processes, thereby furthering our understanding of the system and revealing the possibility that SS~433 is a PeVatron candidate accelerating particles to the $\gtrsim$\,PeV energy range.

The structure of this paper is as follows. In Section~\ref{data reduction}, we describe the data reduction process for the \textit{XMM-Newton} observations and present the imaging and spatially resolved spectral results from these observations. Section~\ref{modeling} introduces leptonic modeling of the nonthermal jets, constrained by the X-ray photon index and intensity profiles, to interpret the multiwavelength spectrum and assess the jet contribution to the $E>100$\,TeV emission. In Section~\ref{discussion}, we discuss the possible contribution of the northern thermal X-ray shell to the UHE emission. Finally, Section~\ref{conclusion} is our conclusion.

\section{Data Analysis and Results} \label{data reduction}
\subsection{Observations and Data Reduction}
In this work, we selected eight \textit{XMM-Newton} \citep{2001A&A...365L...1J} observations of SS~433 region, centered within a $2^{\circ}$ radius to cover the entire jet structure as well as the northern thermal X-ray shell. The details of these observations are listed in Table~\ref{tab:xmmobs}. We performed data reduction using the \textit{XMM-Newton} Science Analysis System (SAS) version 20.0.0 \citep{2004ASPC..314..759G}, in combination with the Extended Source Analysis Software (ESAS) version 0.11.4 to model the extended source backgrounds \citep{2004ApJ...610.1182S}. Each observation was reprocessed using the SAS tasks \texttt{emchain} and \texttt{epchain} to generate calibrated event files for the EPIC-MOS1/MOS2 \citep{2001A&A...365L..27T} and EPIC-PN \citep{2001A&A...365L..18S} cameras, ready for scientific analysis. Time periods with elevated background due to soft proton flares were filtered using the \texttt{mos-filter} and \texttt{pn-filter} tasks with default \texttt{allowsigma}. For observations exhibiting significant residual soft proton contamination after this initial filtering, a stricter sigma-clipping threshold (\texttt{allowsigma=1.5}) was applied. Otherwise, the default parameter (\texttt{allowsigma=3.0}) was retained.

\begin{deluxetable*}{cccccc}
\tabletypesize{\scriptsize}
\tablewidth{0pt} 
\tablecaption{\textit{XMM-Newton} Observations of SS\,433 \label{tab:xmmobs}}
\tablehead{
\colhead{Obs. ID} & \colhead{Obs. date} & \colhead{Duration\,[ks]} & \colhead{Region} & \colhead{Instrument} & \colhead{Offset\,[$^{\prime}$]}
}
\colnumbers
\startdata 
0694870201 & 2012-10-03 & 134.7 & center & EPIC-MOS1/MOS2 & 0.002  \\
0840490101 & 2020-03-24 & 69.1 & eastern-jet1 & EPIC-MOS1/MOS2 & 20.924 \\
0075140401 & 2004-09-30 & 32.5 & eastern-jet2 & EPIC-MOS1/MOS2/PN & 35.621 \\
0075140501 & 2004-10-04 & 31.3 & eastern-jet3 & EPIC-MOS1/MOS2/PN & 61.609 \\
0890800101 & 2022-04-05 & 60.8 & western-jet1 & EPIC-MOS1/MOS2/PN & 25.450 \\
0904520101 & 2023-04-15 & 105.2 & western-jet2 & EPIC-MOS1/MOS2/PN & 41.621 \\
0882560101 & 2022-04-03 & 52.2 & W50-north-east & EPIC-MOS1/MOS2/PN & 22.229 \\
0882560201 & 2021-10-08 & 45.4 & W50-north-west & EPIC-MOS1/MOS2/PN & 22.607 \\
\enddata
\tablecomments{(1)\,\textit{XMM-Newton} Observation ID. (2)\,Observation date. (3)\,The actual observation on-time or duration. (4)\,Pointing region of SS~433 system within a $2^{\circ}$ radius. (5)\,\textit{XMM-Newton} instrument used for data analysis. (6)\,Pointing offset from the central source.}
\end{deluxetable*}

\subsection{Imaging Analysis}
To better characterize the extended spatial structure of the SS~433 system, we employed the \texttt{cheese} task to detect point sources within each detector’s field of view, which were subsequently masked from the analysis. In addition, several residual point sources not detected by \texttt{cheese} were manually excluded. Counts images and exposure maps in different energy bands ($0.4-1.25$\,keV, $1.25-2.0$\,keV, and $2.0-7.2$\,keV) were generated for each observation using \texttt{mos-spectra} and \texttt{pn-spectra}. The corresponding instrumental backgrounds, known as quiescent particle backgrounds (QPBs), were produced with \texttt{mos$\_$back} or \texttt{pn$\_$back} and transformed into sky coordinates using \texttt{rot-im-det-sky}. The final combined, exposure-corrected, background-subtracted, and adaptively smoothed images of the SS~433 region across different energy bands were produced using the \texttt{merge$\_$comp$\_$xmm} and \texttt{adapt$\_$merge} tools. 

\begin{figure*}[ht!]
\plotone{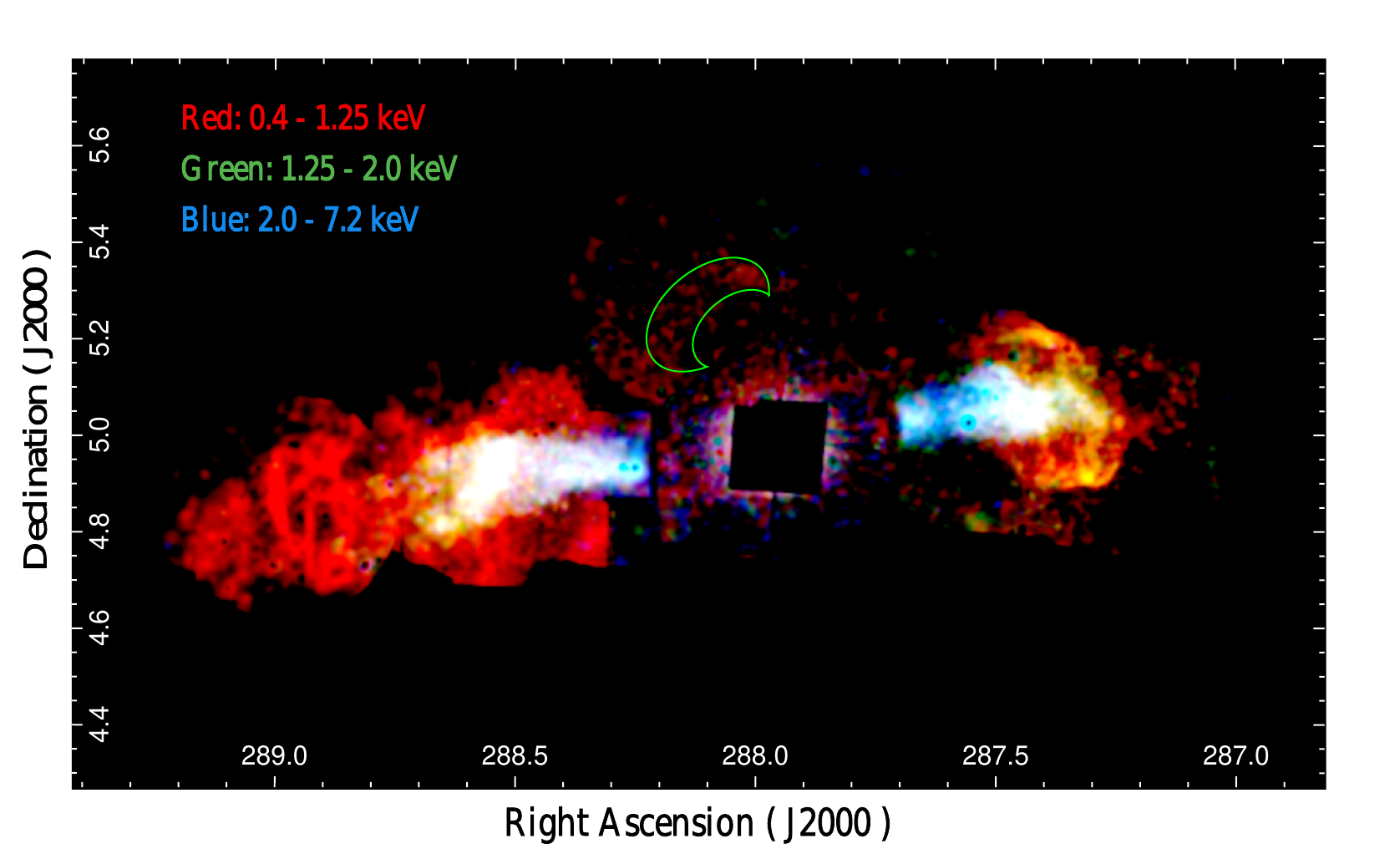}
\caption{Exposure-corrected image of SS~433 on the combined EPIC events with quiescent particle background subtracted, smoothed with a Gaussian kernel of $\sigma=3$ pixels in asinh scale. Red:\,0.4-1.25\,keV. Green:\,1.25-2.0\,keV. Blue:\,2.0-7.2\,keV. The green crescent region outlines the thermal X-ray shell identified by \citet{2024ApJ...975L..28C}.
\label{fig:broad_skymap}}
\end{figure*}

The combined broadband image is displayed in Figure~\ref{fig:broad_skymap}. To avoid pile-up effects from the extremely luminous source, the central CCD of ObsID 0694870201, which was pointed at the compact object, was excluded from the observation. In the $2.0-7.2$\,keV energy band, characterized by nonthermal component, the emission reappears at distances of $18.1^{\prime}$ east and $16.5^{\prime}$ west from the central source and primarily traces the bipolar jets. Localized intensity enhancements, or knots, previously reported by \citet{2007A&A...463..611B} and \citet{2022PASJ...74.1143K} are clearly visible. In the $0.4-1.25$\,keV energy band, dominated by thermal component, the emission appears more extended, filling nearly the entire system and becoming particularly prominent at the jet terminals. The northern thermal X-ray shell, identified as the SNR shell structure \citep{2024ApJ...975L..28C}, can also be seen. The $1.25-2.0$\,keV range is the intermediate energy band, where the emission lines from the \textit{XMM-Newton} instrument become significant in the background spectra.


\subsection{Spatially Resolved Spectral Analysis} \label{jet spectrum}
\begin{figure*}[ht!]
\plotone{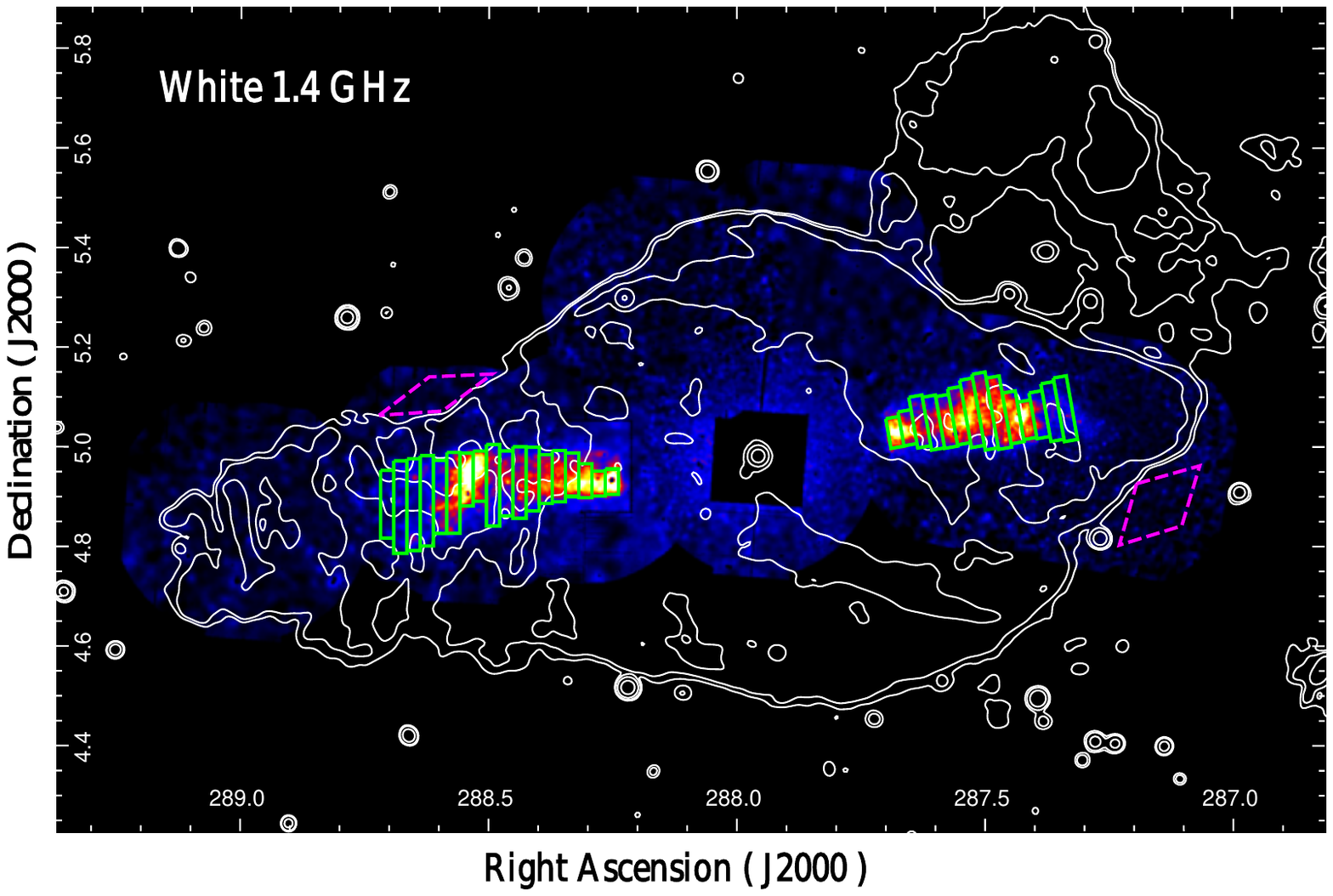}
\caption{Same image as Figure~\ref{fig:broad_skymap}, shown in $2.0-7.2$\,keV band, smoothed with a Gaussian kernel of $\sigma=1.5$ pixels in linear scale. White contour is the 1.4\,GHz VLA observation \citep{1998AJ....116.1842D}. The green boxes delineate the regions used for spectral extraction and the dashed magenta diamonds outside the X-ray and radio emission are used to estimate the sky background.
\label{fig:hard_skymap}}
\end{figure*}

To study the spatial evolution of the nonthermal X-ray emission along both the eastern and western jets, we aim to obtain their intensity and photon index profiles as a function of distance from the central source. To define the transverse boundaries perpendicular to the jet propagation direction, we constructed a fine grid across the $2.0-7.2$\,keV map and extracted counts flux profiles along each transverse direction. When counts flux drops to 1/e of the brightest cell, we consider this to be the boundary of the jet. For the subsequent spectral extraction, the cells within the boundary were merged into elongated rectangular regions oriented perpendicular to the jet axis (green boxes in Figure~\ref{fig:hard_skymap}).

We extracted source spectra from the green rectangular regions and sky background spectra from the magenta dashed diamonds displayed in Figure~\ref{fig:hard_skymap}, with the instrumental background subtracted. The spectra were grouped with \textit{grppha} to ensure at least 30 counts per energy bin, and all spectra were fitted using XSPEC (version 12.13.0c, \citealt{1996ASPC..101...17A}). The sky background spectrum can be modeled as a combination of an unabsorbed thermal component from the Local Hot Bubble ($kT \sim0.1$\,keV); an absorbed, hotter thermal component from the Galactic interstellar medium (ISM, $kT \sim 0.25-0.7$\,keV); and an absorbed power-law component with photon index $\Gamma \sim 1.45$ representing the unresolved cosmic X-ray background from distant extragalactic sources \citep{1997MNRAS.285..449C,2008A&A...478..575K}. We first fitted the eastern and western sky background spectra separately, and then included the best-fit background model in the fits to the source spectra, with the parameters fixed to the best-fit values. The model components and best-fit parameters of the sky background are summarized in Table~\ref{tab:bkg_par} in Appendix~\ref{xspec_spec}. The source spectrum was modeled with either an absorbed power-law (PL) or an absorbed PL plus an \texttt{APEC} thermal component, and the preferred model was selected based on the improvement in $\chi^{2}$ relative to the change in degrees of freedom. The absorption component of all these models is \texttt{tbabs} \citep{2000ApJ...542..914W}. The XMM-Newton spectra of the sky background and representative source regions, together with the best-fitting models folded through the instrumental responses, are shown in Figure~\ref{fig:xspec_spectrum} in Appendix~\ref{xspec_spec}.

\begin{figure*}[ht!]
\plotone{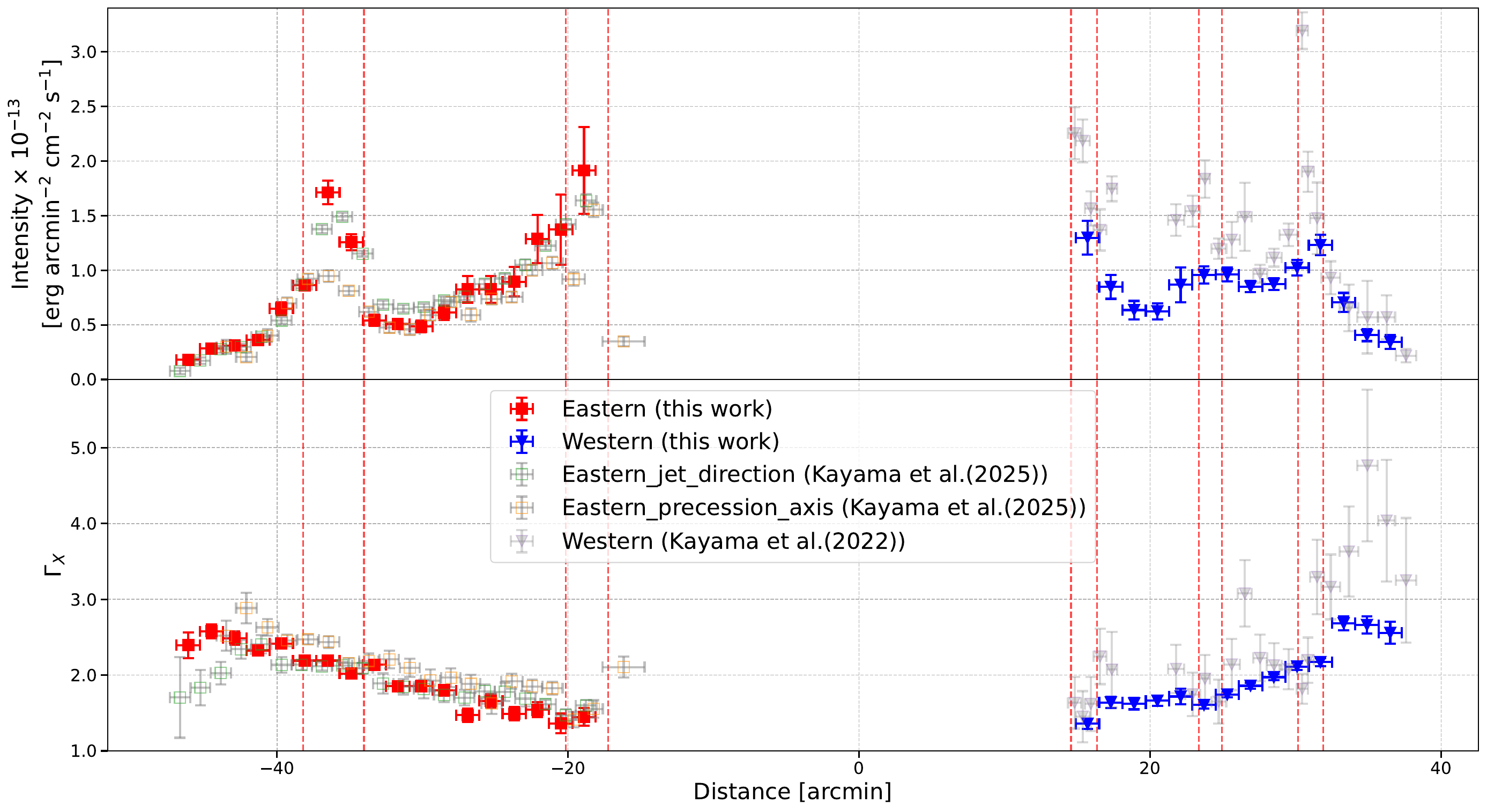}
\caption{Radial profiles of the $1.0-7.0$\,keV nonthermal intensity and photon index, derived from the green extraction regions in Figure~\ref{fig:hard_skymap}. For comparison, the gray data points are taken from \citet{2025PASJ...77..880K}, where one profile was extracted along the jet propagation direction and another along the precession axis for the eastern jet, while the western data are taken from the \citet{2022PASJ...74.1143K}. The vertical dashed lines mark the jet bases and re-brightening knot regions.
\label{fig:profile}}
\end{figure*}

We derived the radial profiles of nonthermal intensity and photon index in $1.0-7.0$\,keV, as shown in Figure~\ref{fig:profile}, with the corresponding numerical values listed in Table~\ref{tab:specresults}. We found that nonthermal emission mainly dominated in the inner jet; with increasing distance from the central black hole, the thermal component became significant. We compared our profiles with those of \citet{2025PASJ...77..880K} and identified several differences. First, our definition of the jet’s transverse boundary differs from theirs. In addition, whereas they constructed two separate profiles with one along the jet axis and another along the precession axis, we derived our profiles by combining all regions exhibiting significant emission. Consequently, the intensity profiles are not directly comparable. For the western jet, the difference may also reflect the background treatment. They adopted the results of \cite{2022PASJ...74.1143K}, which were based on \textit{Chandra} observations. In \cite{2022PASJ...74.1143K}, the source emission fills the \textit{Chandra} field of view, so a local background spectrum from a region free of source emission could not be extracted. Instead, the particle background, the sky background, which consists of the Galactic Ridge X-ray Emission (GRXE), cosmic X-ray background (CXB), and foreground emission, and the source components were fitted simultaneously in the spectra extracted from the source regions, rather than being constrained from a local background spectrum. In contrast, the wider \textit{XMM-Newton} coverage used here allows us to extract local sky background spectra from regions outside the X-ray and radio emission, as shown by the magenta dashed diamonds in Figure~\ref{fig:hard_skymap}. In the resulting profiles, the knot regions show enhanced intensity while the photon index remains relatively soft, resembling the trend seen in the eastern jet.

In addition, when the spatially resolved X-ray spectra are summed, the resulting total spectrum exhibits a slight concave curvature. Although this feature is not statistically significant (see the statistical tests in Appendix~\ref{test_curve}), its origin can be readily understood as a consequence of spatial spectrum mixing. In particular, the superposition of the relatively hard emission from regions close to the jet base and the softer emission at larger distances, especially from the bright re-brightening knots, can naturally produce such a spectral shape.

We use the X-ray profiles and the multiwavelength spectrum to study the origin of the TeV $\gamma$-ray emission and to assess the leptonic contribution to the $E>100$\,TeV emission in Section~\ref{modeling}.

\begin{deluxetable*}{ccccc}
\tabletypesize{\scriptsize}
\tablewidth{0pt}
\tablecaption{Results of spatially resolved \textit{XMM-Newton} spectral analysis \label{tab:specresults}}
\tablehead{
\colhead{Box} & \colhead{distance $z$\,[$^{\prime}$]} & \colhead{$\Gamma_{\rm X}$} & \colhead{Intensity\,[10$^{-14}$ erg arcmin$^{-2}$ cm$^{-2}$ s$^{-1}$]} & \colhead{R\,[$^{\prime}$]}
}
\colnumbers
\startdata 
e1 & 18.1 & 1.45 $\pm$ 0.12 & 19.13 $\pm$ 3.98 & 1.53  \\
e2 & 19.7 & 1.36 $\pm$ 0.13 & 13.73 $\pm$ 3.21 & 1.50 \\
e3 & 21.3 & 1.54 $\pm$ 0.10 & 12.86 $\pm$ 2.20 & 2.04 \\
e4 & 22.9 & 1.49 $\pm$ 0.09 & 8.95 $\pm$ 1.34 & 2.62 \\
e5 & 24.5 & 1.66 $\pm$ 0.09 & 8.25 $\pm$ 1.22 & 3.06 \\
e6 & 26.1 & 1.47 $\pm$ 0.09 & 8.26 $\pm$ 1.22 & 2.98 \\
e7 & 27.7 & 1.80 $\pm$ 0.07 & 6.12 $\pm$ 0.65 & 3.83 \\
e8 & 29.3 & 1.85 $\pm$ 0.07 & 4.85 $\pm$ 0.52 & 4.31 \\
e9 & 30.9 & 1.85 $\pm$ 0.06 & 5.07 $\pm$ 0.43 & 3.91 \\
e10 & 32.5 & 2.14 $\pm$ 0.06 & 5.41 $\pm$ 0.52 & 4.91 \\
e11 & 34.1 & 2.02 $\pm$ 0.04 & 12.57 $\pm$ 0.74 & 3.03 \\
e12 & 35.7 & 2.19 $\pm$ 0.04 & 17.12 $\pm$ 1.09 & 3.18 \\
e13 & 37.3 & 2.19 $\pm$ 0.04 & 8.63 $\pm$ 0.42 & 4.87 \\
e14 & 38.9 & 2.42 $\pm$ 0.06 & 6.49 $\pm$ 0.58 & 4.40 \\
e15 & 40.5 & 2.33 $\pm$ 0.06 & 3.62 $\pm$ 0.28 & 5.44 \\
e16 & 42.1 & 2.49 $\pm$ 0.09 & 3.10 $\pm$ 0.33 & 5.52 \\
e17 & 43.7 & 2.58 $\pm$ 0.09 & 2.83 $\pm$ 0.31 & 5.58 \\
e18 & 45.3 & 2.39 $\pm$ 0.17 & 1.81 $\pm$ 0.39 & 4.08 \\
\hline
w1 & 16.5 & 1.36 $\pm$ 0.07 & 12.97 $\pm$ 1.55 & 1.82 \\
w2 & 18.1 & 1.64 $\pm$ 0.08 & 8.47 $\pm$ 1.08 & 2.07 \\
w3 & 19.7 & 1.63 $\pm$ 0.08 & 6.35 $\pm$ 0.84 & 3.10 \\
w4 & 21.3 & 1.66 $\pm$ 0.07 & 6.23 $\pm$ 0.74 & 3.23 \\
w5 & 22.9 & 1.72 $\pm$ 0.10 & 8.67 $\pm$ 1.59 & 3.16 \\
w6 & 24.5 & 1.61 $\pm$ 0.05 & 9.58 $\pm$ 0.80 & 3.43 \\
w7 & 26.1 & 1.74 $\pm$ 0.04 & 9.61 $\pm$ 0.63 & 4.11 \\
w8 & 27.7 & 1.86 $\pm$ 0.04 & 8.50 $\pm$ 0.52 & 4.63 \\
w9 & 29.3 & 1.97 $\pm$ 0.04 & 8.74 $\pm$ 0.54 & 4.35 \\
w10 & 30.9 & 2.11 $\pm$ 0.05 & 10.23 $\pm$ 0.70 & 3.10 \\
w11 & 32.5 & 2.17 $\pm$ 0.05 & 12.30 $\pm$ 0.93 & 2.29 \\
w12 & 34.1 & 2.68 $\pm$ 0.09 & 7.05 $\pm$ 0.87 & 2.80 \\
w13 & 35.7 & 2.66 $\pm$ 0.11 & 4.05 $\pm$ 0.55 & 3.53 \\
w14 & 37.3 & 2.56 $\pm$ 0.14 & 3.43 $\pm$ 0.65 & 3.89 \\
\enddata
\tablecomments{(1)\,Regions used for the spatially resolved spectral analysis shown in Figure~\ref{fig:hard_skymap}; the 18 rows above correspond to the eastern jet, while the 14 rows below correspond to the western jet. (2)\,Projected distance of each box from the central source. (3)\,Photon index of the nonthermal emission extracted from each box. (4)\,Intensity of the nonthermal emission extracted from each box. (5)\,Transverse radius of each box.}
\end{deluxetable*}

\section{Modeling} \label{modeling}

\subsection{Steady-state multi-component fitting} \label{steady-state-fitting}

The variability of SS~433 jets is low. This implies that although particles in different regions of the jet continuously flow in and out and undergo radiative cooling, the particle number and energy spectrum within each region remain in dynamic equilibrium, maintaining stable emission across the jet, at least within the observational period. Therefore, we first assume a steady-state electron population in each spatial bin and test whether the resulting synchrotron and IC emission can reproduce the observed X-ray and $\gamma$-ray data.

In the calculation, the electron spectrum in each bin is modeled as a broken power-law. The minimum Lorentz factor is fixed at 1, and the maximum Lorentz factors for the eastern and western jets are denoted as $\gamma_{\rm max,e}$ and $\gamma_{\rm max,w}$, noting that the maximum Lorentz factors are not important for the fitting. The break Lorentz factor of the electron spectrum is set to $10^7$ and the low-energy spectral index $s_1$ is assumed to be $-2$. These two parameters would not affect the fitting of X-ray and TeV $\gamma$-ray data, but may influence the total energy in the electron spectrum, which is not the main focus of this study. On the other hand, the high-energy spectral index $s_2$ can be inferred from the observed X-ray spectrum directly (i.e., $s_2=2\Gamma_{\rm X}-1$). The main free parameters are normalization of the spectrum (i.e., total electron energy) and the magnetic field. For the moment, we simply assume that the magnetic field within each jet is constant, with $B_e$ for the eastern jet and $B_w$ for the western jet being independent of the distance from the central compact object $z$. For the X-ray data fitting, the electron spectrum normalization and the magnetic field are coupled. To reproduce the observed intensity in each bin, different choices of the magnetic field profile lead to different required electron normalizations within the region. Consequently, the predicted IC emission also changes accordingly. We may test the leptonic origin of LHAASO observed UHE $\gamma$-ray emission, by comparing the sum of the IC radiation from each spatial bin with the $\gamma$-ray data. Here, we consider two external photon fields for the IC radiation: the cosmic microwave background (CMB) and far-infrared (FIR) photons. The CMB is a blackbody spectrum with a temperature of $2.7\,{\rm K}$, while the FIR radiation is modeled as a graybody spectrum with a temperature of $30\,{\rm K}$ and an energy density of $10^{-12}\,{\rm erg\,cm^{-3}}$ according to the prediction for the location of SS~433 by the interstellar radiation model given by \citet{Popescu2017}. Higher-temperature component of the background radiation is not important for TeV $\gamma$-ray photons due to the Klein-Nishina (KN) suppression.

\begin{figure*}[ht!]
\centering
\includegraphics[width=0.55\textwidth]{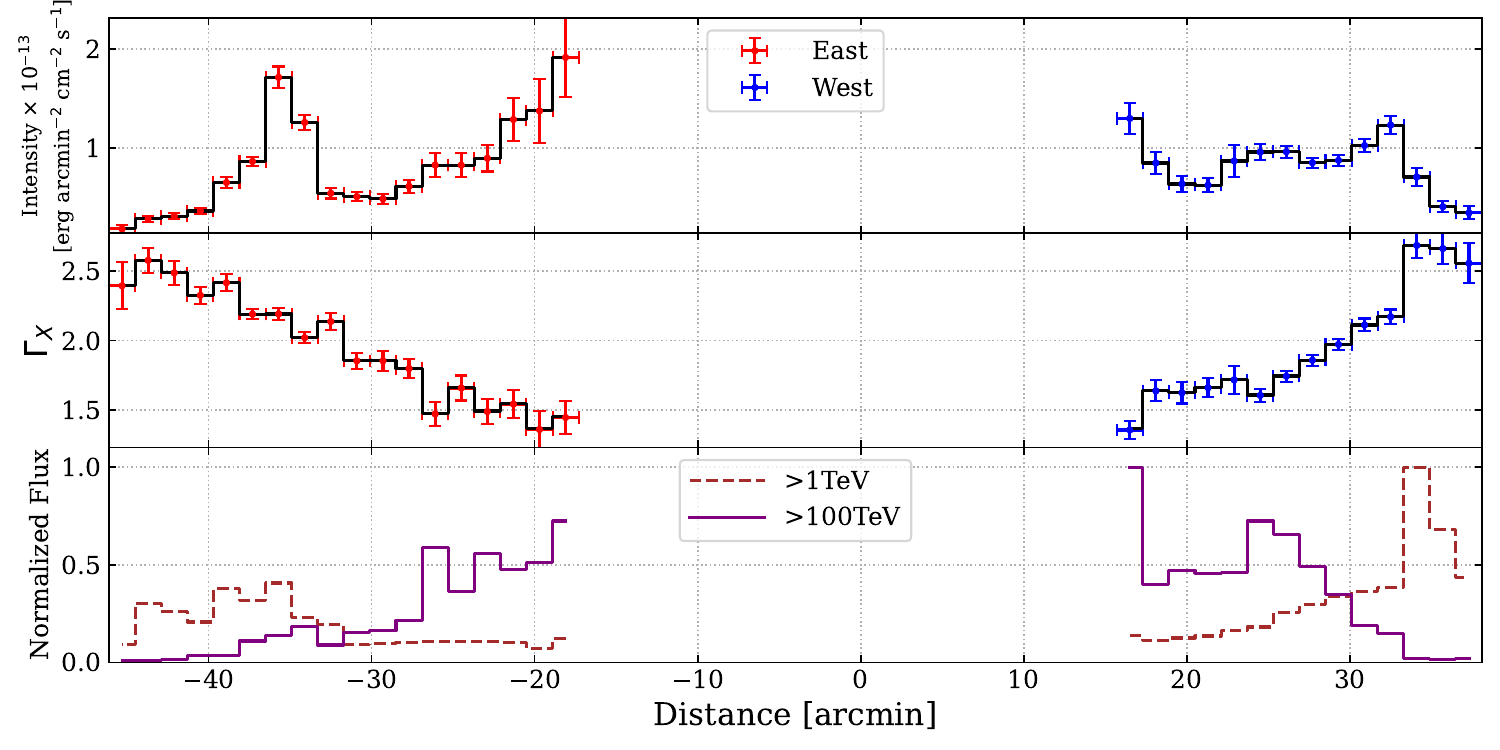}
\includegraphics[width=0.35\textwidth]{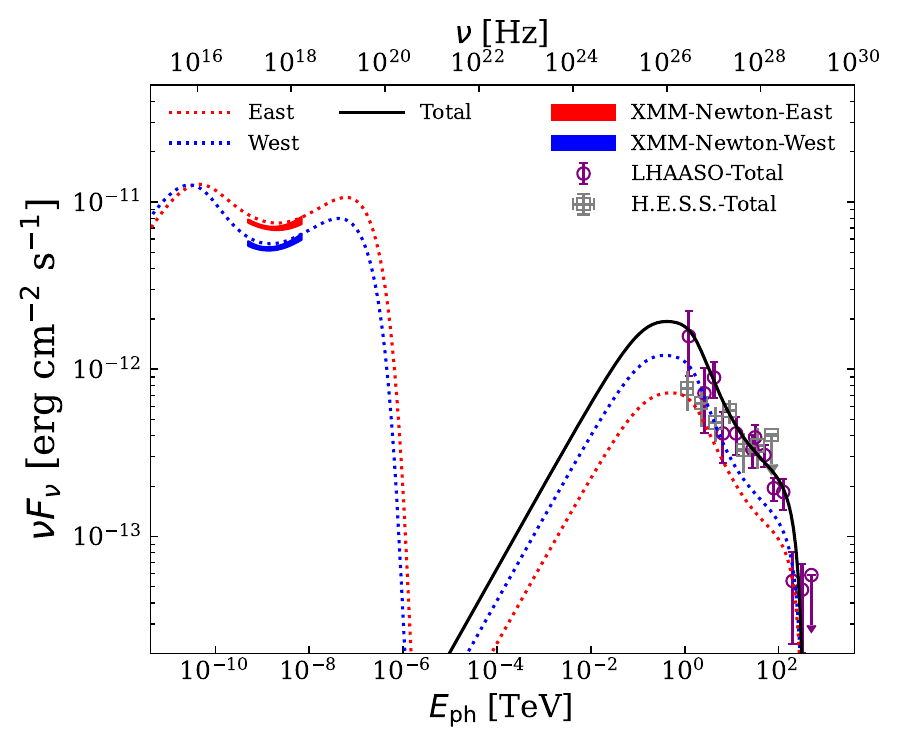}
\caption{Profile and SED of steady-state fitting. The upper and middle panel of the left figure are the fitting results of intensity and photon index profiles. The lower panel is the normalized TeV flux profile along the jets. The normalizations are $4.25\times10^{-13}\,{\rm erg\,cm^{-2}\,s^{-1}}$ for above 1\,TeV and $1.49\times10^{-14}\,{\rm erg\,cm^{-2}\,s^{-1}}$ for above 100\,TeV. The right figure is the modeling of SED. The red and blue dashed lines are the summed spectra of the eastern and western jet respectively. The black solid line is the total IC radiation.}
\label{fig:steady_profile&sed}
\end{figure*}

According to \citet{10.1093/nsr/nwaf496}, IC radiation from a single-zone leptonic component can adequately explain the data up to $20-30$\,TeV, while higher-energy emission is difficult to reproduce. Similarly, in this work, we aim to first optimize the spectral fitting to the data below 20\,TeV, and then examine how well the model can reproduce the spectrum at hundreds of TeV. With suitable choices of the normalization and magnetic field, this benchmark model can reproduce the X-ray intensity and photon index profiles as well as the $\gamma$-ray spectrum from a few TeV to hundreds of TeV, as shown in Figure~\ref{fig:steady_profile&sed}. The best-fit magnetic fields are $B_e=18.8\,\mu$G and $B_w=14.5\,\mu$G for the eastern and western jets, respectively. The maximum electron energies are assumed to be $\gamma_{\rm max,e}=\gamma_{\rm max,w}=7.5\times10^8$. In this simplified picture, $\gtrsim 100\,$TeV emission is mainly contributed by the inner jet bins, where the electron spectrum is harder, and the lower-TeV emission is dominated by larger distances along the jet regions. The contributions of the individual spatial bins to the multiwavelength spectral energy distributions (SEDs) of the eastern and western jets are shown in Figure~\ref{fig:sed_steady} in Appendix~\ref{sed_steady}. We emphasize, however, that this solution serves mainly as a baseline case. It does not demonstrate that the UHE emission is uniquely explained by electrons in the jet, nor does it establish that such a constant magnetic field profile is physically preferred.

\subsection{Jet profile and magnetic field configuration}

We further discuss the effects of different magnetic field configurations. The magnetic field profile of the SS~433 jets is likely to be scale- and region-dependent, while its evolution in strength along the jets remains poorly constrained. Radio polarimetry on subarcsecond-to-arcsecond scales generally finds the projected 
magnetic field to be aligned with the precessing jet ridge line, although ALMA observations indicate a transverse field close to the central source that becomes longitudinal farther downstream, near the region where discrete ejecta begin to interact \citep{stirling2004, roberts2008, blundell2018}. On the much larger scale of the eastern W50 ear, Faraday-rotation measurements reveal loop-like or helical structures together with magnetic fields shaped by the bow and terminal shocks \citep{farnes2017,sakemi2018}. Most relevant to the nonthermal X-ray-emitting region considered here, IXPE detected a highly polarized signal from the head of the eastern jet, implying a substantial ordered magnetic-field component parallel to the bulk flow \citep{kaaret2024}. Such a longitudinal field may represent a poloidal component carried outward by the jet, but it may also be produced locally by velocity shear, compression, or interactions between discrete ejecta \citep{lopezmiralles2026}. Therefore, the existing observations do not determine a unique global field geometry or magnetic-field strength profile. We here examine two idealized flux-conserving limits for an expanding jet. Under the assumption of an incompressible, sub-relativistic jet, conservation of magnetic flux along the jet would lead to a magnetic field configuration of $B\propto R^{-1}$ if the toroidal component of the magnetic field is dominant, or $B\propto R^{-2}$ if poloidal component is dominant. 

\begin{figure*}[ht!]
\centering
\includegraphics[width=0.9\textwidth]{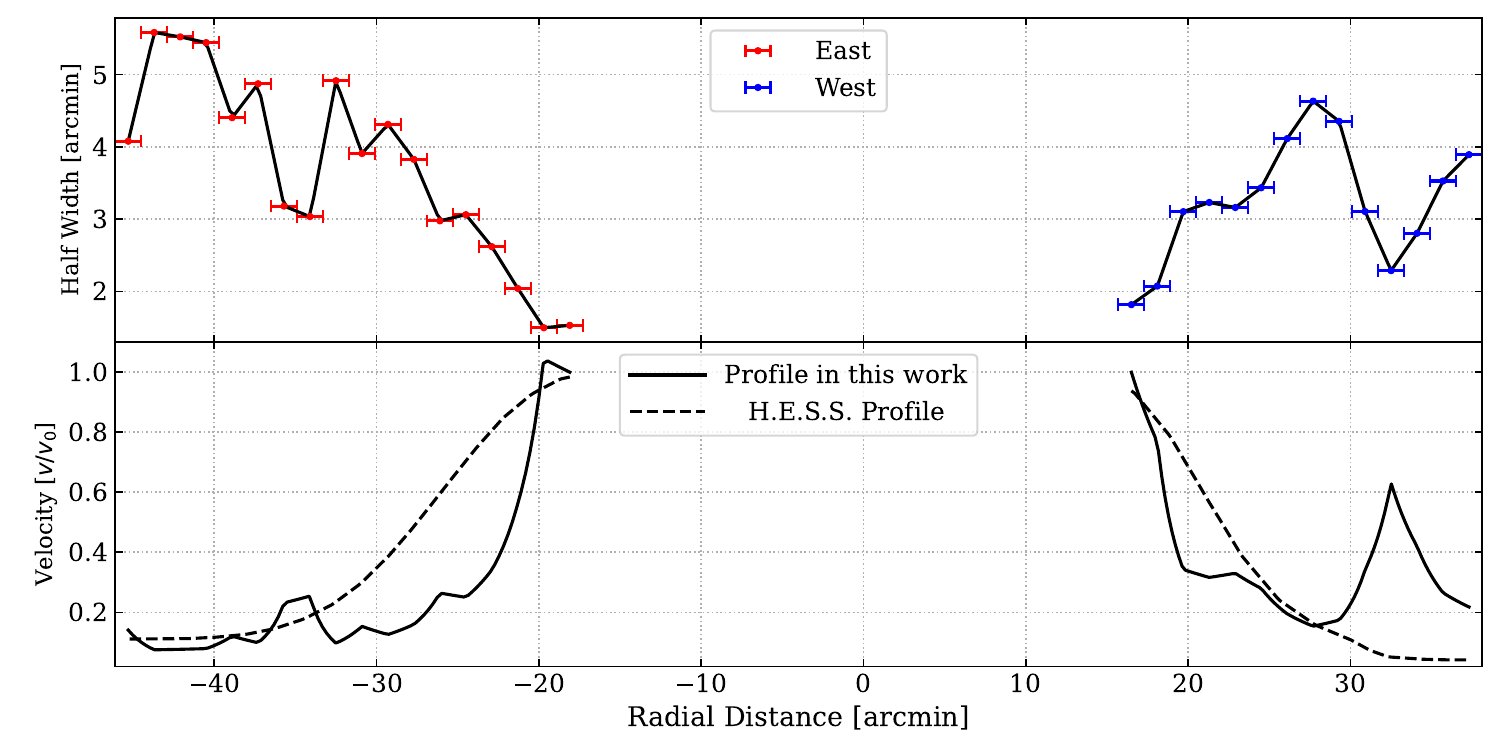}
\caption{Profiles of the jet transverse radius and velocity. In the upper panel, the red and blue points represent the half-widths derived from the jet boundaries identified in Section~\ref{jet spectrum}. The lower panel presents the velocity profile obtained under the assumptions that the jet is axisymmetric and incompressible (solid lines). For comparison, H.E.S.S-measured velocity profile is displayed as dashed line.}
\label{fig:jet_profile}
\end{figure*}

We first consider the velocity profile adopted in the H.E.S.S. analysis (as indicated by the dashed line in the lower panel of Figure~\ref{fig:jet_profile}) \citep{2024Sci...383..402H}. Under the assumption of an incompressible jet, the velocity profile $v(z)$ corresponds to a transverse radius profile $R(z)$\footnote{In fact, $v(z)$ shown in H.E.S.S. analysis was derived from $R(z)$. The latter was based on the \textit{ROSAT} observation of the jet's morphology, but was not explicitly shown in \citet{2024Sci...383..402H}. We derived $R(z)$ based on $v(z)$ provided in the paper.}. The magnetic field profile, $B(z)$, can be also derived in both toroidal and poloidal magnetic field cases if the magnetic flux is conserved within the jet. Under this profile, the model produces softer $\gamma$-ray spectra and fails to reproduce the $\gamma$-data above 30\,TeV satisfactorily. This is because when the magnetic field decays rapidly with $z$, the inner jet regions must be more strongly magnetized, with $B_{e,0}=45.2\,\mu {\rm G}, B_{w,0}=56.3\,\mu {\rm G}$ at the jet base in the case of $B\propto R^{-1}$, or $B_{e,0}=120\,\mu {\rm G},B_{w,0}=245\,\mu {\rm G}$ in the case of $B\propto R^{-2}$. The strong magnetic field at the inner jet region suppresses the IC radiation, leading to an overall softer $\gamma$-ray spectrum and insufficient flux at the highest energies.

\begin{figure*}[ht!]
\centering
\includegraphics[width=0.45\textwidth]{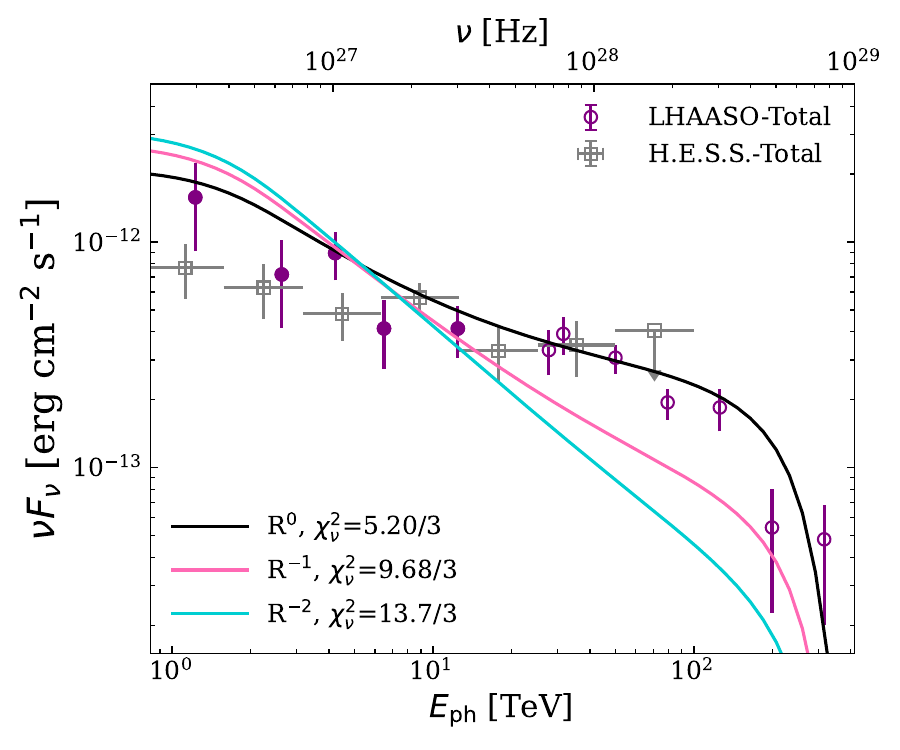}
\includegraphics[width=0.45\textwidth]{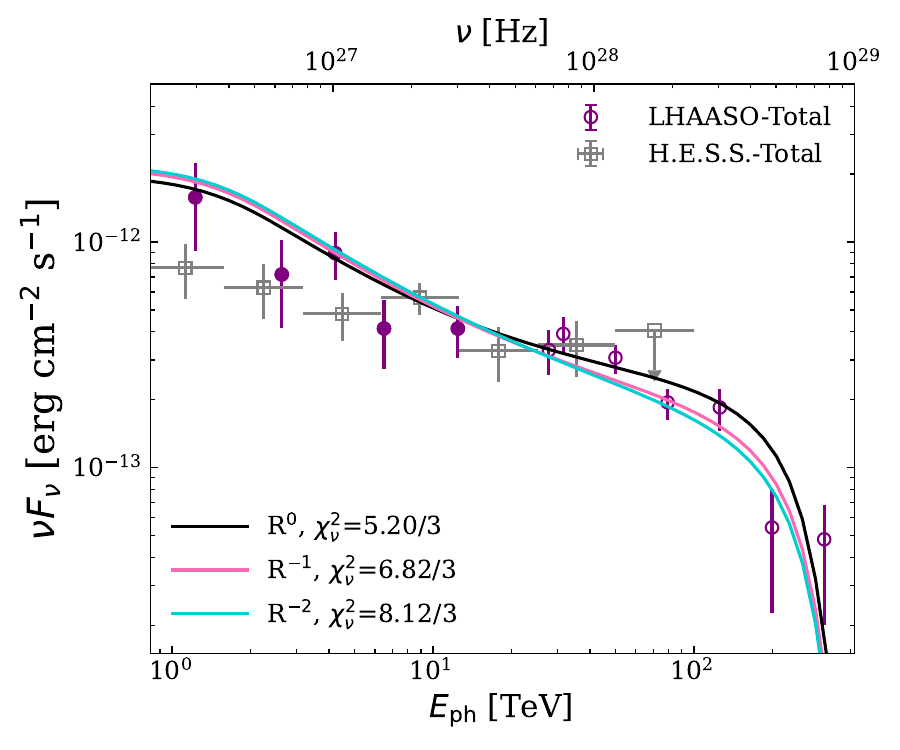}
\caption{TeV SEDs with different configurations of magnetic field based on H.E.S.S-based velocity profile (left) and velocity profile inferred from the nonthermal X-ray morphology in this work (right).}
\label{fig:SED_comparison}
\end{figure*}


We note, however, that the jet profile is not uniquely determined observationally, because different X-ray bands may trace different physical components of the jet. In particular, the profile used in the H.E.S.S. study was inferred from softer X-ray ($0.9-2.0$\,keV) contours, which may include a non-negligible contribution from thermal plasma, especially at large distances. Given that the nonthermal X-ray component is most directly relevant to the synchrotron/IC-emitting electrons considered here, we consider an alternative transverse radius profile $R(z)$ based on the half-width of each spatial bin in the $2.0-7.2$\,keV observed by \textit{XMM-Newton}, which mostly reflect the distribution of nonthermal electrons. More specifically, we start from the contour of the nonthermal jet on the sky map, as shown in Figure~\ref{fig:hard_skymap}. The half-width of each spatial bin can be approximately regarded as the local radius of the jet. The upper panel of Figure~\ref{fig:jet_profile} shows the half-width (i.e., jet radius) at different distances. To derive the velocity profile of the jet, we assume that the jet is axisymmetric and incompressible following \citet{2024Sci...383..402H}. In such an instance, mass flux is conserved within the jet, implying that the velocity is inversely proportional to the cross-sectional area. The resulting velocity profile is shown as the solid line in the lower panel of Figure~\ref{fig:jet_profile}. Applying such jet radius and velocity profiles, the magnetic fields are found to be $B_{e,0}=52.2\,\mu$G and $B_{w,0}=27.0\,\mu$G for $B\propto R^{-1}$, while $B_{e,0}=162\,\mu$G and $B_{w,0}=54.4\,\mu$G for $B\propto R^{-2}$ to explain the $\gamma$-ray emission. The right panel of Figure~\ref{fig:SED_comparison} shows the results based on the profile inferred from the nonthermal X-ray morphology. We see that the mismatch between the model prediction and the LHAASO UHE data becomes less visually striking. Nevertheless, the predicted flux around 100\,TeV is still lower than the observed data as indicated by the chi-squared test. In this sense, the preference for an additional UHE component is weakened, but not fully removed, when the nonthermal X-ray-based profile is used.

It is also worth noting that X-ray data alone do not impose a strong constraint on the maximum electron energy. The highest photon energy observed by \textit{XMM-Newton} is about $\sim7$\,keV, which can be produced via synchrotron radiation by electrons with energies $E_{e}\approx 40~{\rm TeV} (E_{\rm syn}/7~{\rm keV})^{1/2} (B/50~\mu {\rm G})^{-1/2}$. Even considering the \textit{NuSTAR} observation of hard X-ray emission up to 30\,keV from the inner jet region \citep{2022ApJ...935..163S}, the maximum electron energy inferred from X-ray observations may still be much lower than the energy required for producing $>100\,$TeV photons, i.e., $\sim 300-400\,{\rm TeV}$. Therefore, X-ray observations by themselves cannot establish whether the highest-energy $\gamma$-ray emission is produced by the same electron population inside the jets. Moreover, if the TeV $\gamma$-ray emission contains a significant hadronic contribution, the required contribution from IC emission of relativistic electrons would be reduced. For a fixed synchrotron X-ray flux, this would allow a stronger magnetic field and a correspondingly smaller number of relativistic electrons in the jets. Therefore, the obtained magnetic field may be interpreted as lower-limit estimate for given jet profiles and magnetic-field configurations.

\subsection{Propagation and evolution of electrons}

\begin{figure*}[ht!]
\centering
\includegraphics[width=0.9\textwidth]{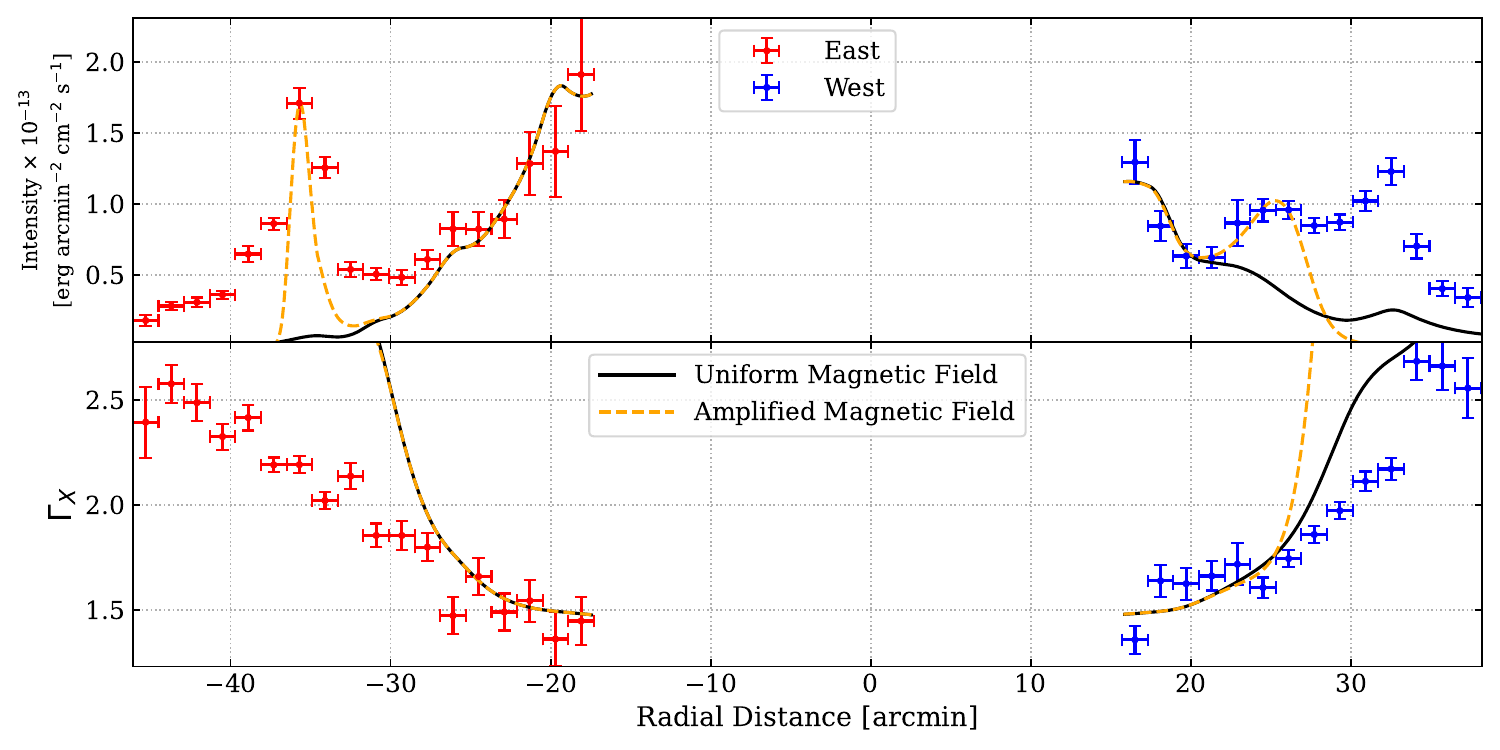}
\caption{X-ray intensity and photon index profiles of the evolution model. The black solid line shows the case where electrons are injected at the base of the jet and then propagate outward, evolving along the jet. The orange dashed line represents a scenario in which there is a sudden enhancement of the magnetic field within the jet.}
\label{fig:evo}
\end{figure*}

In reality, electron distribution within a jet is governed by injection, transport and cooling. In the scenario envisaged by \citet{2024Sci...383..402H}, electrons are accelerated at the jet base and then advected outward along the jet axis. In the absence of further reacceleration, electrons gradually cool due to radiative and adiabatic losses. Following this scenario, we assume that electrons are injected at the bases of both the eastern and western jets with injection luminosity $L_{\rm inj,e}$ and $L_{\rm inj,w}$ respectively. The injected electron spectrum follows power-law distribution $Q(\gamma)=L_{\rm inj}/(m_e c^2 \int \gamma^{1-s}d\gamma) \gamma^{-s}$ with the index $s_e$ and $s_w$ for the eastern and western jet respectively. The evolution of electron population can be described as
\begin{equation}
    \frac{\partial N(\gamma,z,t)}{\partial t}=-\frac{\partial}{\partial \gamma}[\dot{\gamma}(\gamma,z,t)N(\gamma,z,t)] + Q(\gamma)\delta(z=z_0),
\end{equation}
where $t=\int_{z_0}^z dz/v(z)$ denotes the propagation time from the jet base. The term $\dot{\gamma}(\gamma,z,t)=\frac{4}{3}\frac{\sigma_T}{m_e c^2}[u_B(z)+F_{\rm KN}(\gamma)(u_{\rm CMB}+u_{\rm FIR})]\gamma^2$ represents radiative energy loss rate. The last term corresponds to the injection term, which occurs only at the base of the jet. 

Here, the magnetic field is assumed to be uniform with $B_e=B_w=20\mu G$. The velocity at the jet base is $v_{e,0}=v_{w,0}=0.26c$. The maximum Lorentz factor electron is set to $\gamma_{\rm max,e}=\gamma_{\rm max,w}=1.0\times 10^9$. The fitting result is shown as the black solid line in Figure~\ref{fig:evo}. Starting from injection, and without any additional re-acceleration or further injections, the electrons continuously cool to lower energies. Therefore the intensity gradually decreases while the photon index steadily increases (i.e., the spectrum becomes softer). As the jet propagates to larger distances, its velocity decreases due to the increase in cross-sectional area. As a result, the high-energy electrons within the flow cool completely before reaching the outer regions. Therefore, under the assumption of a single electron injection process, the X-ray profile at large distances cannot be explained. In addition, both the eastern and western jets exhibit re-brightening features, which also cannot be accounted for in this scenario.

We further test whether local magnetic-field amplification at the knot positions can account for the intensity enhancements. We artificially enhance the magnetic field strength at the knot position. More specifically, magnetic field enhancements are introduced as a simple Gaussian profile $\Delta B(z) = B_{\rm en} e^{-(z-z_p)^2/2\sigma^2_z}$ peaked around $z_p=37^\prime$ in the eastern jet and $z_p=28\,^\prime$ in the western jet. The enhancement amplitudes and region sizes are set to $B_{\rm en}=4.5B_e,\ \sigma_z=2^\prime$ and $B_{\rm en}=1.5B_w,\ \sigma_z=2.5^\prime$ for the eastern and western jets, respectively. As expected, the intensity in these regions increases and may more or less reproduce the local intensity profiles (displayed as the dashed orange line in Figure~\ref{fig:evo}). However, the enhanced magnetic field leads to more severe electron cooling, and subsequently lower intensity and softer spectra at larger distances. Therefore, magnetic field amplification alone cannot explain the  re-brightening (or the knot structures) observed in the profiles.

\section{Discussion} \label{discussion}

In addition to the contribution of IC radiation of electrons accelerated at jet bases, the $E>100$\,TeV $\gamma$-rays detected in SS~433 may also originate from electrons accelerated at other sites or through hadronic processes. Notably, the extended UHE emission is spatially coincident with a thermal X-ray shell north of SS~433 reported by \citet{2024ApJ...975L..28C}, which also aligns with radio continuum of the W50 nebula and an H\textsc{i} cloud \citep{2020NatAs...4.1177L}. Motivated by these spatial correlations, we investigate whether the UHE emission can originate from IC radiation of electrons accelerated at the SNR shock, or via $pp$-collision of accelerated protons interacting with the atomic cloud.

\begin{figure}[ht!]
\centering
\includegraphics[width=0.9\columnwidth]{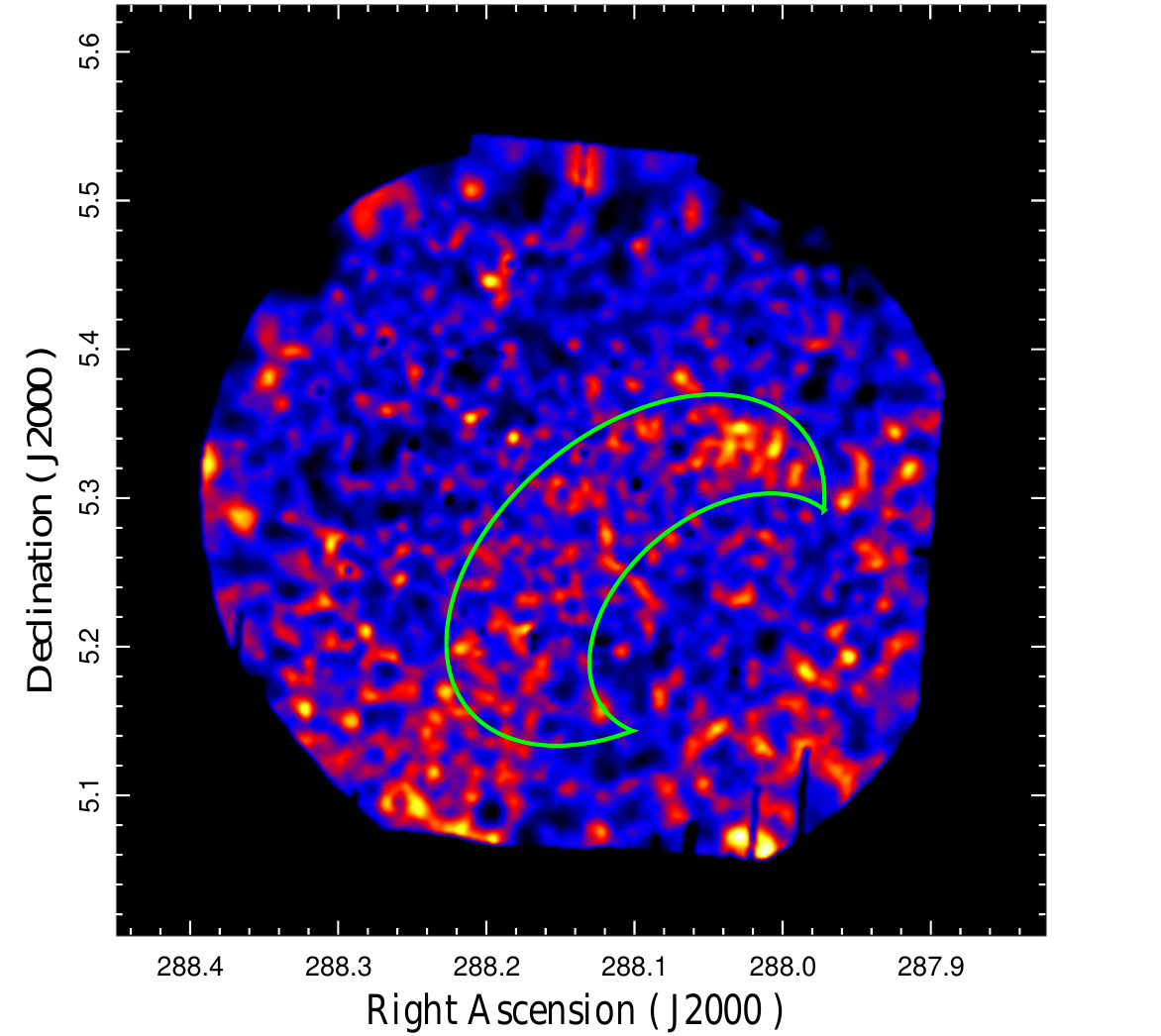}
\caption{Exposure-corrected and quiescent-particle-background-subtracted image of the thermal shell north of SS~433 in $0.4-1.25$\,keV, smoothed with a Gaussian kernel of $\sigma=1.5$ pixels in linear scale. The green region, following \citet{2024ApJ...975L..28C}, is used to analyze the emission from the northern shell.
\label{fig:north}}
\end{figure}

We therefore search for nonthermal X-ray emission in the northern thermal X-ray shell, which would help to constrain critical quantities such as the particle acceleration efficiency and magnetic field of the SNR. Figure~\ref{fig:north} displays the northern shell in the $0.4-1.25$\,keV energy band, with the source spectrum extracted from the green region. The background spectrum was extracted from a source-free region located outside both the X-ray and radio shells in the same observation, and the background was modeled following \citet{2024ApJ...975L..28C}. We have successfully reproduced the source spectral results reported in \citet{2024ApJ...975L..28C} with an absorbed \texttt{VNEI} model, which is widely used to describe non-equilibrium ionization collisional plasma in SNRs \citep{2001ApJ...548..820B}, yielding thermal emission characterized by $kT \sim 1$\,keV. To assess the potential presence of a nonthermal synchrotron emission driven by shock-accelerated electrons in the SNR, we add an additional absorbed power-law component to the source model. However, the nonthermal contribution is poorly constrained, allowing us to derive only an upper limit on its flux. Given this X-ray upper limit, we perform a multiwavelength SED analysis combined with LHAASO data. We follow the scenario proposed by \citet{10.1093/nsr/nwaf496}, where emission below a few tens of TeV is from the two jets, whereas the emission above a few tens of TeV is from the thermal shell. To account for the emission around and above 100\,TeV, we introduce a second component modeled by a power-law spectrum of electrons with a cutoff at $E_{\rm cut} \approx 300$\,TeV. In this leptonic scenario, the absence of nonthermal X-ray detection constrains the magnetic field to $\lesssim 9~\mu$G.

\begin{figure}[ht!]
\centering
\includegraphics[width=0.9\columnwidth]{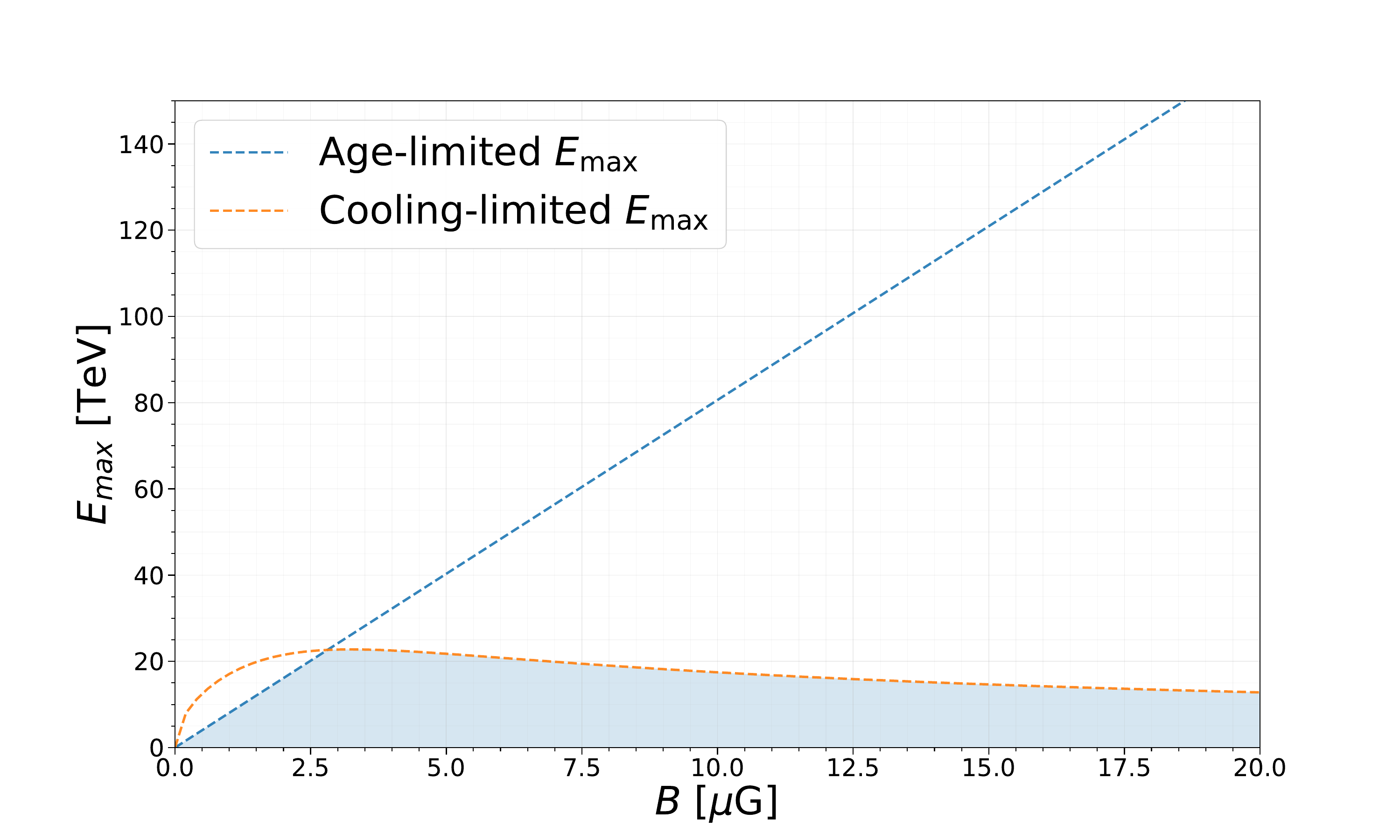}
\caption{The maximum electron energy attainable at the SNR shock as a function of magnetic field strength. The blue dashed line shows the age-limited maximum energy, while the orange dashed line shows the cooling-limited maximum energy. The shaded region indicates the allowed maximum electron energy, defined by the lower of the two limits.}
\label{fig:Emax}
\end{figure}

\citet{2024ApJ...975L..28C} estimated an SNR age of $20-30$\,kyr from the shell properties and derived a shock velocity of $u_{\rm s} \approx 923\,\rm km\,s^{-1}$ from the temperature of the 1\,keV underionized plasma. With these parameters, we can estimate the maximum electron energy that can be produced by the SNR. In the framework of the diffusive shock acceleration (DSA, \citealt{1978ApJ...221L..29B,1987PhR...154....1B}) mechanism, the acceleration timescale can be estimated by \citep{2001JPhG...27.1589K,2007Ap&SS.309..119R}
\begin{equation}
t_{\rm acc} \approx \frac{3D}{{u_{\rm s}}^2} = \eta (\frac{r_{\rm g}}{c}) (\frac{c}{u_{\rm s}})^2 \simeq \eta (\frac{E}{eBc}) (\frac{c}{u_{\rm s}})^2
\end{equation}
where $D=\eta r_{\rm g} c/3$ is the spatial diffusion coefficient of the particle, $r_{\rm g}$ is the particle's Larmor radius, $c$ is the speed of light, $B$ is the magnetic field near the shock front, and $\eta=D/D_{\rm Bohm} \geq 1$ represents the deviation of particle diffusion from the Bohm diffusion, with $\eta=1$ corresponding to the Bohm limit. \citet{2006A&A...453..387P} inferred diffusion coefficients of $\sim 1-10$ times the Bohm diffusion coefficient for the highest-energy electrons in young SNRs. Since the SNR shell considered here is substantially older and has a much slower shock, a relatively large diffusion coefficient is plausible. We therefore adopt $\eta=10$, corresponding to the upper end of the range inferred for young SNRs, as an optimistic value for the maximum electron energy attainable in this evolved SNR shell. We then estimate the maximum energy by equating the acceleration timescale to the SNR age or the radiative cooling (synchrotron and IC) timescale. The obtained $E_{\rm max}$ is determined by the lower of these two, yielding its dependence on the magnetic field strength as shown in Figure~\ref{fig:Emax}. We find that the achievable electron energy peaks at only $\sim 20$\,TeV, which is significantly lower than the required energy (at least 300\,TeV) of the primary electrons responsible for the observed $>100$\,TeV $\gamma$-rays. 

Therefore, electrons accelerated by the SNR shock are unlikely to explain the UHE emission. A hadronic origin in which the UHE emission is produced by interactions of escaping PeV protons with the H\textsc{i} cloud may be feasible. These PeV protons might be accelerated at earlier evolution stage of the SNR when the shock velocity and the magnetic field strength were higher, although in the hadronic scenario escaping protons may originate from SS~433 as well \citep{2020ApJ...904..188K,10.1093/nsr/nwaf496, Peretti2025}.

\section{Conclusion} \label{conclusion}
In this work, we analyzed \textit{XMM-Newton} observations covering the bipolar jets and the thermal X-ray shell north of SS~433, and derived spatially resolved X-ray intensity profile and photon-index profile along both jets. The profiles show an overall spectral softening with increasing distance from the central source, together with pronounced knots and re-brightening features on both sides.

As a phenomenological benchmark, we perform a bin-by-bin leptonic modeling constrained by the spatially resolved X-ray measurements. The spatial-resolved X-ray photon index directly reflects the evolution of electron spectrum along the jet. At the innermost jet region, the inferred electron spectral index is as hard as approximately $E^{-1.8}$. Under a uniform magnetic field configuration, this hard electron component at the inner jet can reproduce the $\gamma$-ray spectrum at UHE regime measured by LHAASO. However, once a more realistic magnetic field configuration with conservation of magnetic flux is considered (such as the one suggested by \citealt{2024Sci...383..402H}), the magnetic field strength at the inner jet region would be significantly higher than that at the outer jet region, resulting in suppression of the IC flux from the inner jet region. Consequently, the overall $\gamma$-ray spectrum becomes soft and cannot match the measured spectrum well. We emphasize that the level of the mismatch depends on the employed jet's transverse radius profile $R(z)$. We considered an alternative transverse radius profile based on nonthermal X-ray morphology of the jet obtained in this work, which is regarded as another plausible description of the jet geometry. The latter profile alleviates the discrepancy, but does not remove the difficulty of reproducing the UHE $\gamma$-ray spectrum. The preference for an additional UHE component is therefore reduced, although not fully eliminated.

It is also important to note that the current X-ray data do not by themselves provide a strong constraint on the maximum electron energy in the jets. The highest photon energy detected by \textit{XMM-Newton} is only $\sim 7$\,keV, and in a relatively strong magnetic field such X-ray photons can be produced by electrons with energies still below 100\,TeV. By contrast, producing $\gamma$-rays above 100\,TeV through inverse-Compton scattering generally requires electrons with energies of at least $\sim 300-400$\,TeV. Therefore, the present X-ray observations alone do not demonstrate that electrons in the jets indeed reach such energies and hence do not provide decisive support for a leptonic origin of the UHE emission yet, even under the case of a uniform magnetic field configuration.

In addition, the modeling considering transport of electrons within jets shows that a single injection site at the jet base cannot account for the observed X-ray intensity and photon index profiles along the jets. Even when local amplification of magnetic field strength is introduced, the re-brightening features and their downstream emission still cannot be explained. We therefore conclude that, the X-ray data favor additional particle injection and/or re-acceleration along the jet.

To put it shortly, X-ray observations suggest an additional spectral component to account for the UHE $\gamma$-ray emission from the SS~433 region. However, the origin of this component is still uncertain at the moment. In the future, UHE $\gamma$-ray observations by high angular resolution instruments such as LACT \citep{2024icrc.confE.808Z,2025ChPhC..49c5001Z}, ASTRI-Mini \citep{2022JHEAp..35...52S} and CTA \citep{2011ExA....32..193A} would be crucial for localizing the production site of the UHE emission from SS~433 and revealing its origin.

\section*{Acknowledgments}
This study is based on observations obtained with \textit{XMM-Newton}, an ESA science mission with instruments and contributions directly funded by ESA Member States and NASA. The authors acknowledge the support of National Natural Science Foundation of China under the grant 12593852 and the Basic Research Program of Jiangsu under grant No.~BK20250059.

\appendix

\setcounter{figure}{0}
\renewcommand{\thefigure}{A\arabic{figure}}
\renewcommand{\theHfigure}{A\arabic{figure}}

\setcounter{table}{0}
\renewcommand{\thetable}{A\arabic{table}}
\renewcommand{\theHtable}{A\arabic{table}}

\section{X-ray Spectra with Folded Models} \label{xspec_spec}
\begin{figure*}[ht!]
\centering
\begin{minipage}{0.49\textwidth}
\centering
\includegraphics[width=\linewidth]{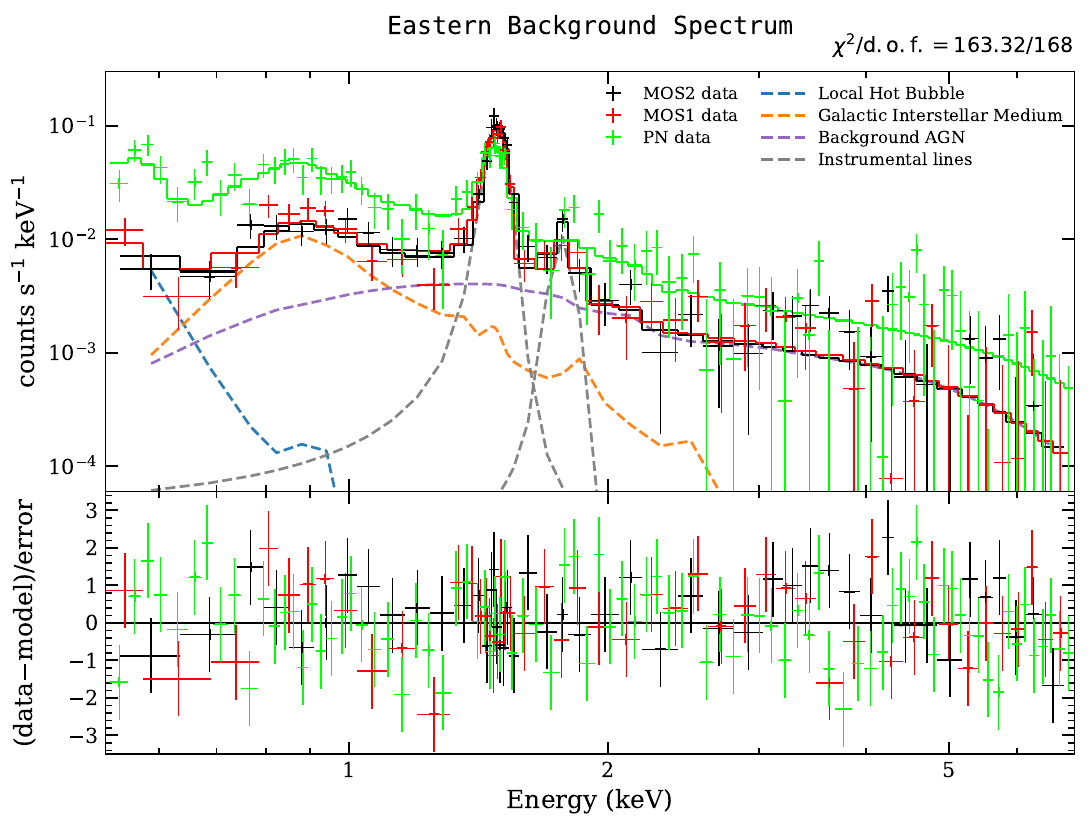}\\[0.8em]
\includegraphics[width=\linewidth]{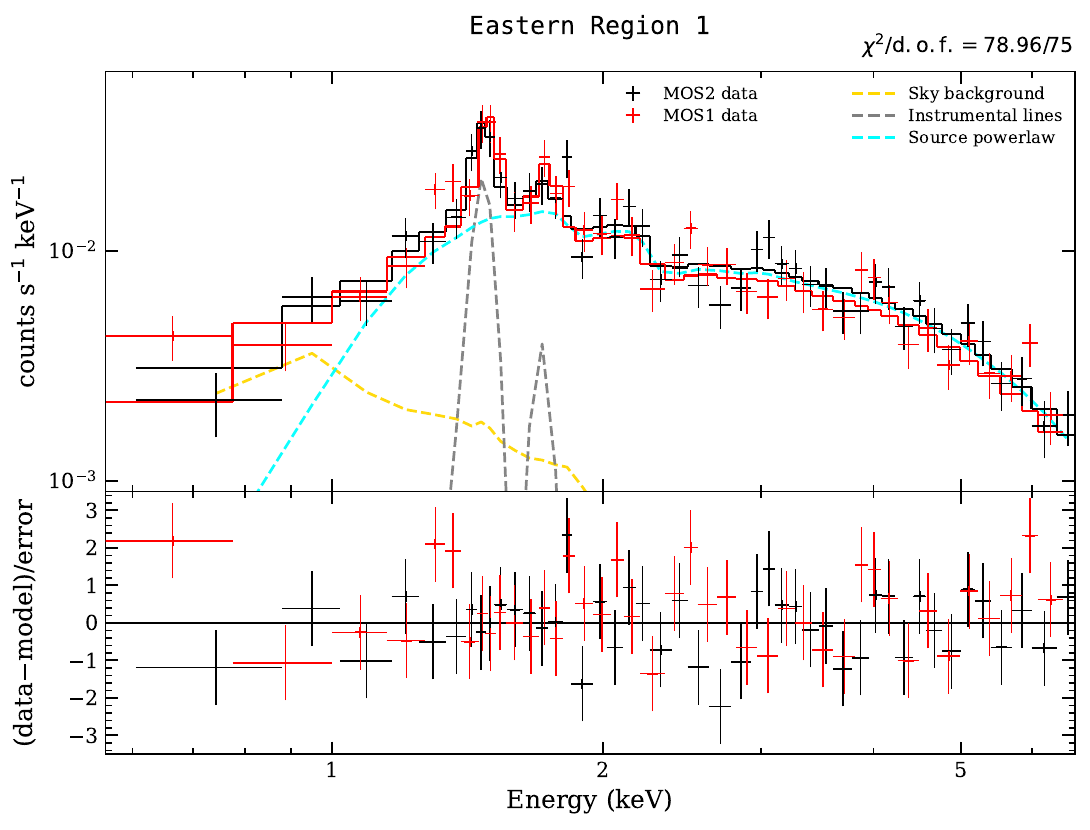}\\[0.8em]
\includegraphics[width=\linewidth]{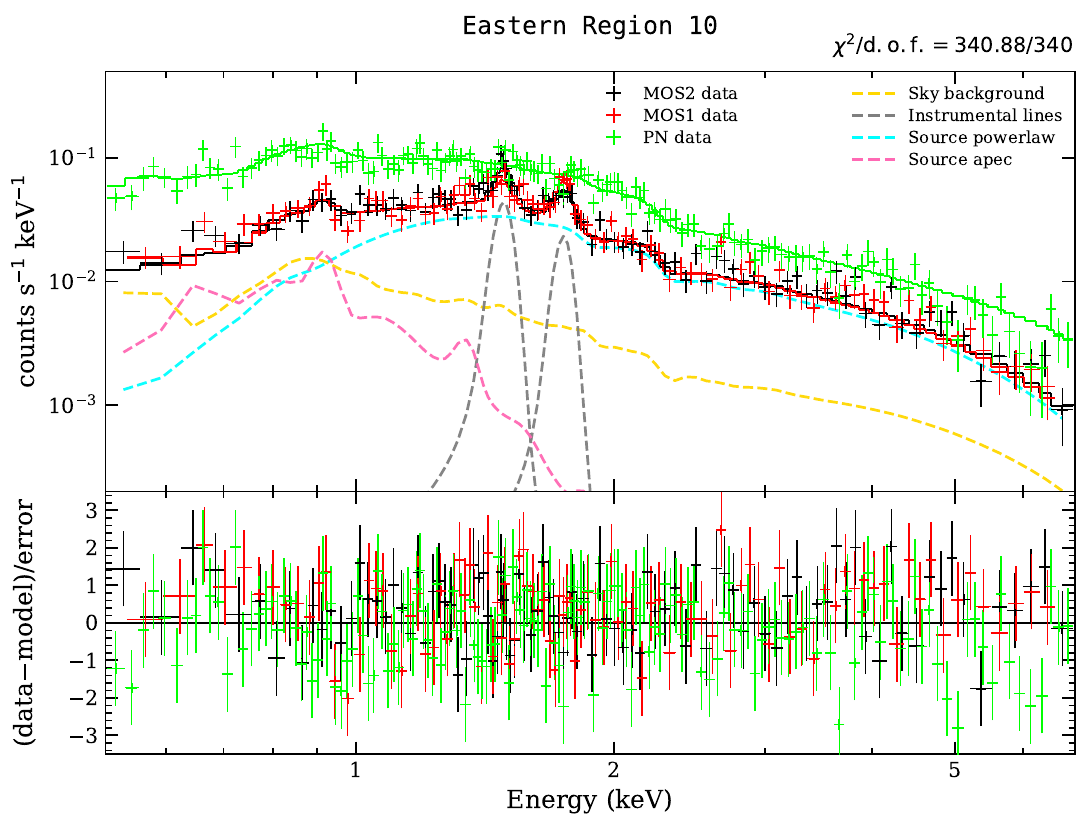}
\end{minipage}
\hfill
\begin{minipage}{0.49\textwidth}
\centering
\includegraphics[width=\linewidth]{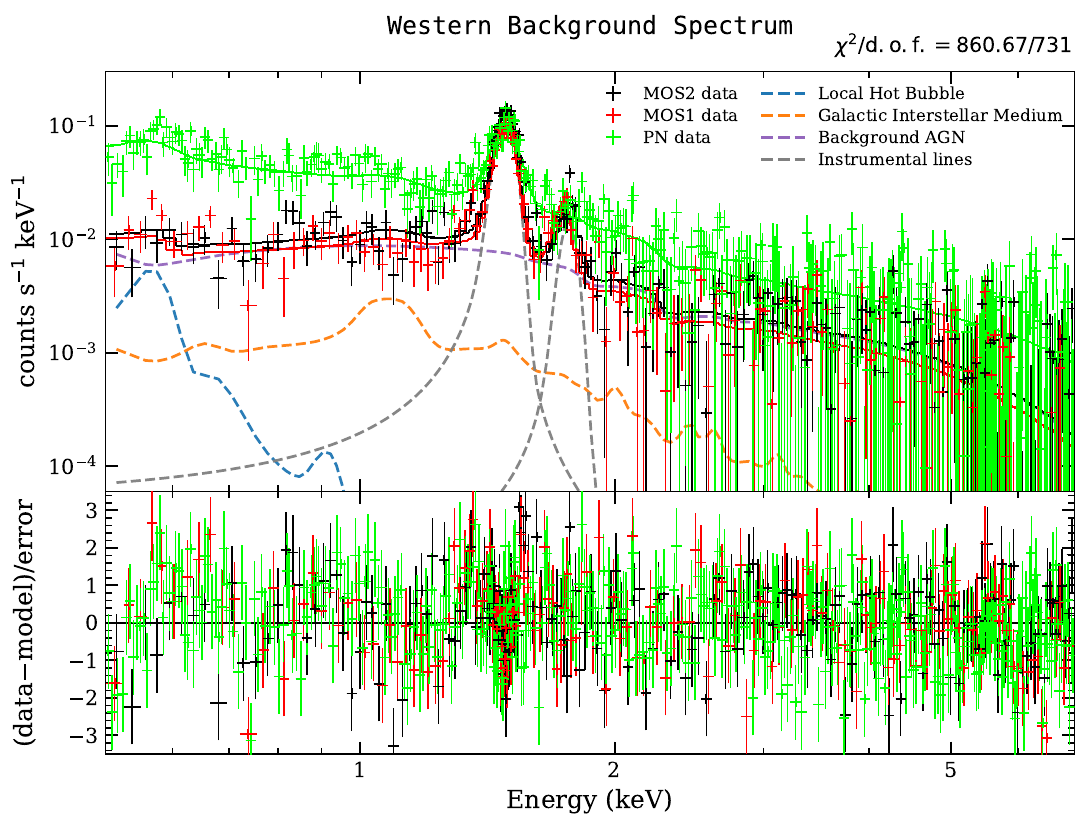}\\[0.8em]
\includegraphics[width=\linewidth]{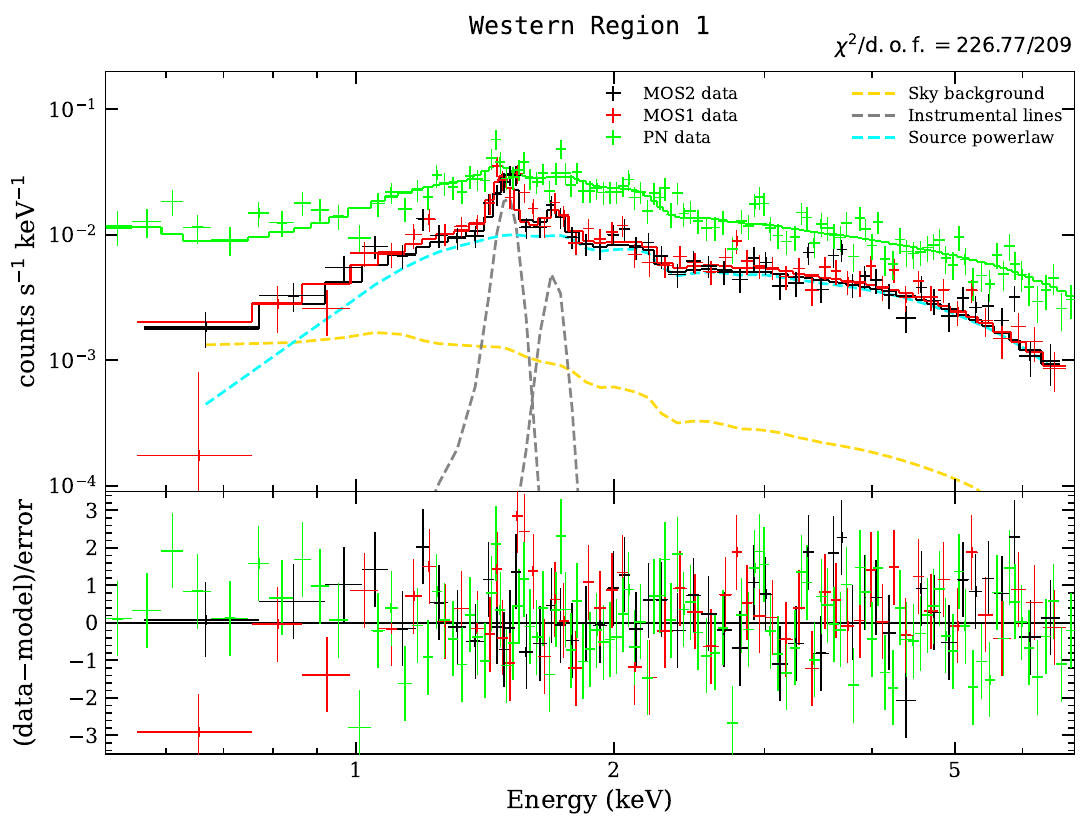}\\[0.8em]
\includegraphics[width=\linewidth]{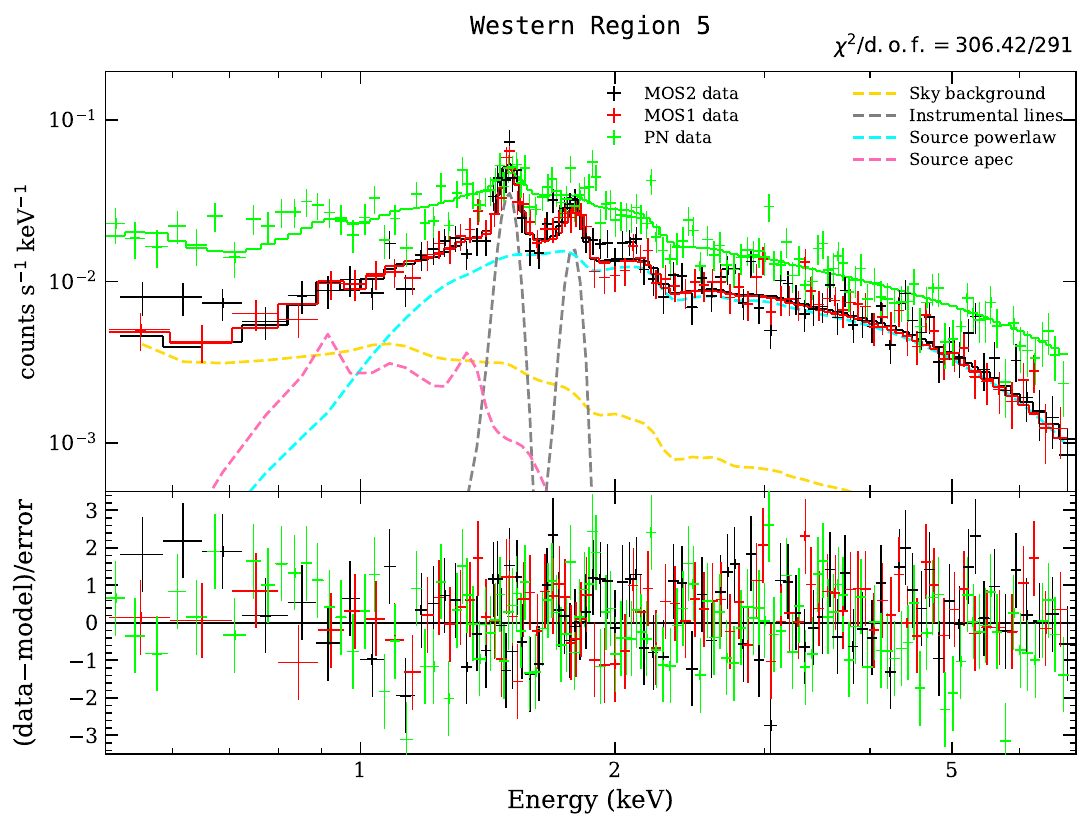}
\end{minipage}
\caption{\textit{XMM-Newton} spectra and best-fitting models folded through the instrumental responses for the background and source regions. The left panels show the eastern background region and source regions e1 and e10, while the right panels show the western background region and source regions w1 and w5. The corresponding nonthermal spectral fitting results for these source regions are reported in Table~\ref{tab:specresults}. These source regions were selected to sample two characteristic spectral regimes, in which e1 and w1 are dominated by nonthermal emission, while a thermal component becomes significant in e10 and w5. In each panel, the data points are from the EPIC MOS2, MOS1, and PN cameras. The dashed curves show the individual folded MOS2 model components, and the solid curves show the total model for each camera.}
\label{fig:xspec_spectrum}
\end{figure*}

We show the \textit{XMM-Newton} spectra for the background and representative source regions in Figure~\ref{fig:xspec_spectrum}, together with the best-fitting models folded through the instrumental responses. For reference, the best-fitting parameters of the sky background model components are summarized in Table~\ref{tab:bkg_par}.

\begin{deluxetable*}{ccccc}
\tabletypesize{\scriptsize}
\tablewidth{0pt}
\tablecaption{Best-fit spectral model parameters for the sky background \label{tab:bkg_par}}
\tablehead{
\colhead{Physical Component} & \colhead{Model component} & \colhead{Parameter} & \colhead{East} & \colhead{West}
}
\startdata
\multirow{2}{*}{Local Hot Bubble} & \multirow{2}{*}{apec} & $kT$ & 0.1, fixed & 0.1, fixed \\
& & norm & $(6.44\pm1.18)\times10^{-6}$ & $(2.42\pm0.52)\times10^{-6}$ \\
\hline
Galactic absorption for sky background & tbabs & $N_{\rm H}$ & $(0.35\pm0.17)\times10^{22}$ & 0 \\
\hline
\multirow{2}{*}{Galactic interstellar medium (ISM)} & \multirow{2}{*}{apec} & $kT$ & $0.77\pm0.06$ & $1.85\pm0.64$ \\
& & norm & $(2.10\pm1.08)\times10^{-6}$ & $(5.30\pm3.84)\times10^{-7}$ \\
\hline
\multirow{2}{*}{Extragalactic background AGNs} & \multirow{2}{*}{power-law} & photon index & 1.45, fixed & 1.45, fixed \\
& & norm & $(1.39\pm0.12)\times10^{-6}$ & $(1.12\pm0.09)\times10^{-6}$ \\
\enddata
\tablecomments{$kT$ is in units of keV. The power-law normalization is in units of photons keV$^{-1}$ cm$^{-2}$ s$^{-1}$ arcmin$^{-2}$ at 1 keV. The apec normalization is in units of cm$^{-5}$ arcmin$^{-2}$. $N_{\rm H}$ is in units of $\rm cm^{-2}$. The errors correspond to 1$\sigma$ uncertainties.}
\end{deluxetable*}

\onecolumngrid

\section{Statistical Tests for Curvature in the Summed X-ray Spectra} \label{test_curve}
\begin{figure*}[ht!]
\centering
\includegraphics[width=0.9\textwidth]{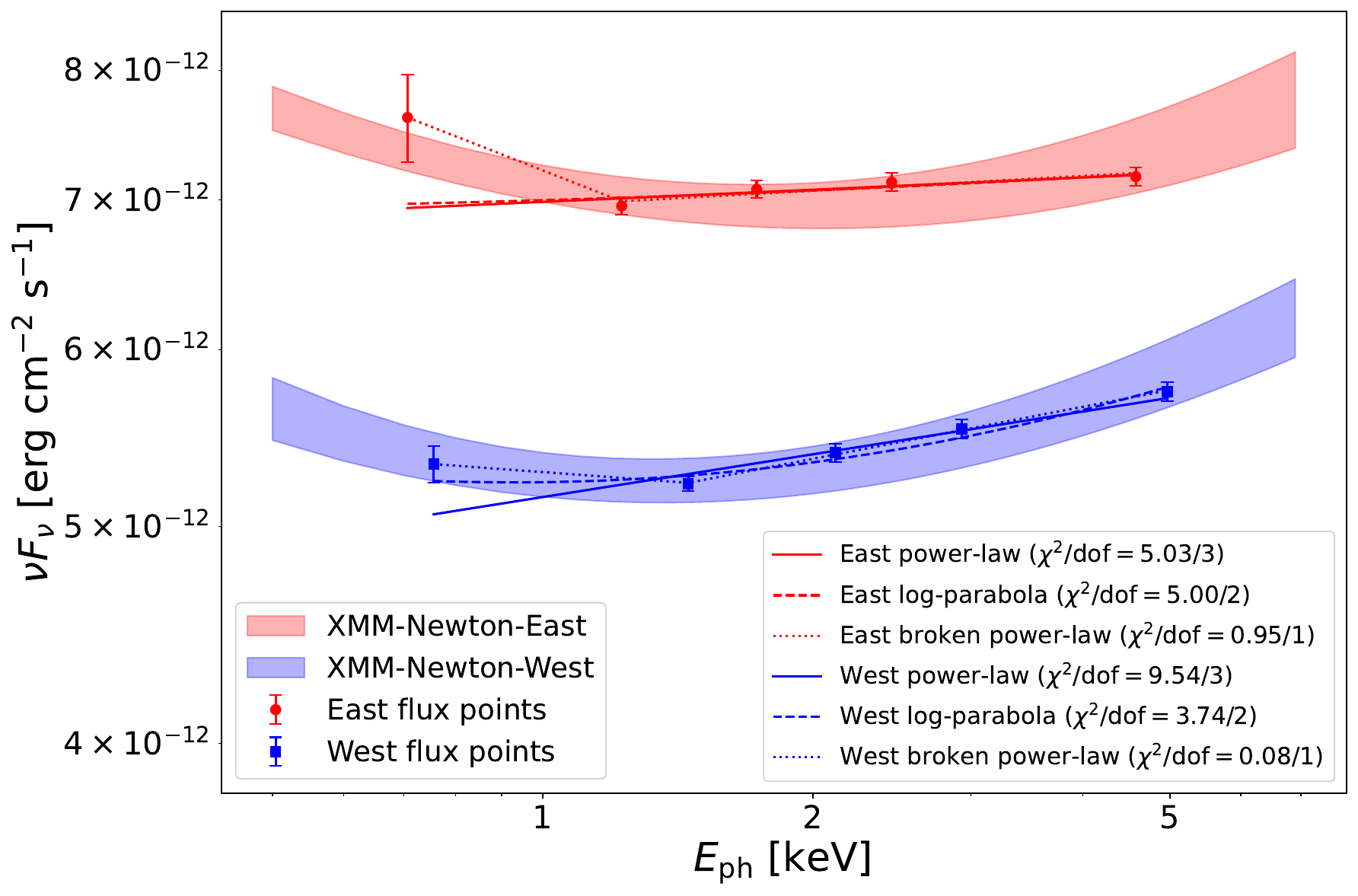}
\caption{Tests for spectral curvature in the X-ray spectra of the eastern and western jets. The light red and blue shaded regions represent the summed spectra of the eastern and western jets, respectively, and correspond to the X-ray spectra shown in the right panel of Figure~\ref{fig:steady_profile&sed}. The red and blue data points were obtained by summing the source flux contributions from all spatial bins within each of the five energy intervals. Solid, dashed, and dotted lines show the best-fitting power-law, log-parabola, and broken power-law models, respectively. The corresponding $\chi^2/ \rm d.o.f.$ values are listed in the legend.}
\label{fig:test_curve}
\end{figure*}

To quantify the significance of the possible upward curvature in the summed X-ray spectra, we fitted the summed spectral flux points from the eastern and western jet regions with power-law, log-parabola, and broken power-law models, as shown in Figure~\ref{fig:test_curve}. For the eastern region, the power-law, log-parabola and broken power-law models yield $\chi^2/\rm{d.o.f.}=5.03/3$, $5.00/2$, and $0.95/1$, respectively. Defining $\Delta{\rm AIC}={\rm AIC}_{\rm model}-{\rm AIC}_{\rm PL}$, we obtain $\Delta{\rm AIC}=1.97$ for the log-parabola model and $-0.08$ for the broken power-law model. For the western region, the corresponding fit statistics are $\chi^2/\rm{d.o.f.}=9.54/3$, $3.74/2$, and $0.08/1$, with $\Delta{\rm AIC}=-3.80$ and $-5.46$ for the log-parabola and broken models, respectively. These results indicate that the eastern spectrum shows no preference for a curved model, whereas the western spectrum shows a modest preference for spectral curvature. However, the improvement is not sufficiently strong to establish significant curvature in either jet region.

\begin{figure*}[ht!]
\centering
\includegraphics[width=0.9\textwidth]{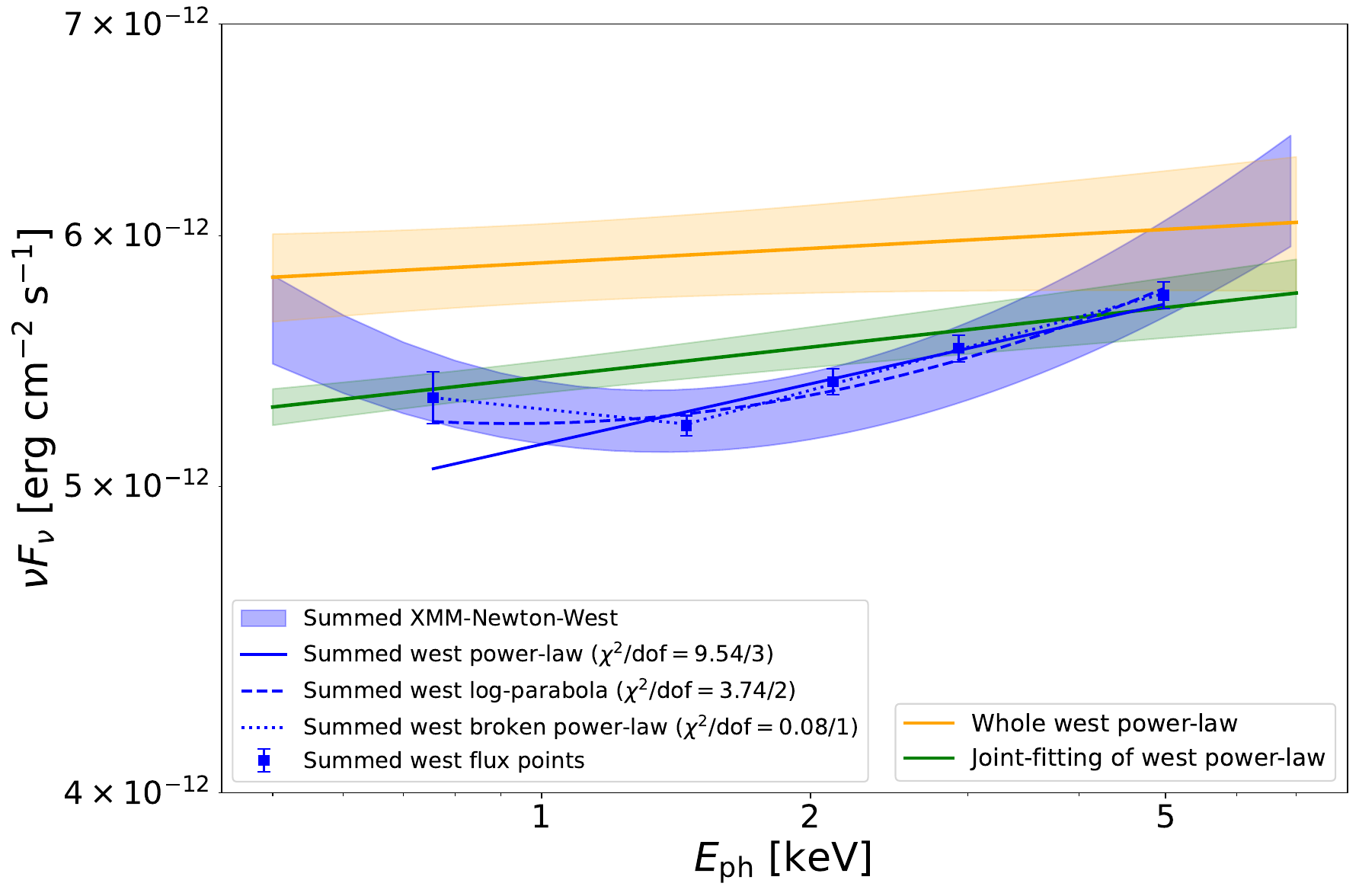}
\caption{Comparison of the summed western jet spectrum with that extracted from the entire western jet and with the result of a joint fit to the spatially resolved spectra. The blue band, data points, and solid, dashed, and dotted lines are identical to those shown in Figure~\ref{fig:test_curve}. The three lines represent the best-fitting power-law, log-parabola, and broken power-law models, respectively, for the five data points. The orange band represents the power-law fit to the spectrum extracted directly from a single polygonal region encompassing the entire western jet, while the green band shows the result of the joint fit to the spatially resolved western jet spectra, with the column density and the photon index of the source power-law component tied across all spatial bins.}
\label{fig:west_test_curve}
\end{figure*}

To investigate the origin of the slight upward curvature in the summed X-ray spectrum, we performed two additional tests for the western jet. The western jet is fully covered by a single observation, making it well suited for direct spectral extraction from a single polygon encompassing the entire jet, whereas the eastern jet was observed in two separate pointings. We therefore extracted and fitted the spectrum of the entire western jet. In addition, we jointly fitted the spectra from all western spatial bins while tying the absorbing column density and the photon index of the source power-law component across each bin. In both tests, an additional thermal component was included to account for possible thermal emission from W50. The results are shown in Figure~\ref{fig:west_test_curve}. These two tests yield approximately consistent spectral shapes, suggesting that the curvature arises primarily from spatial mixing among spectra with independently fitted column densities and photon indices.

\section{Spatially Resolved SEDs in the Steady-State Multi-component Model} \label{sed_steady}
\begin{figure*}[ht!]
\centering
\includegraphics[width=0.9\textwidth]{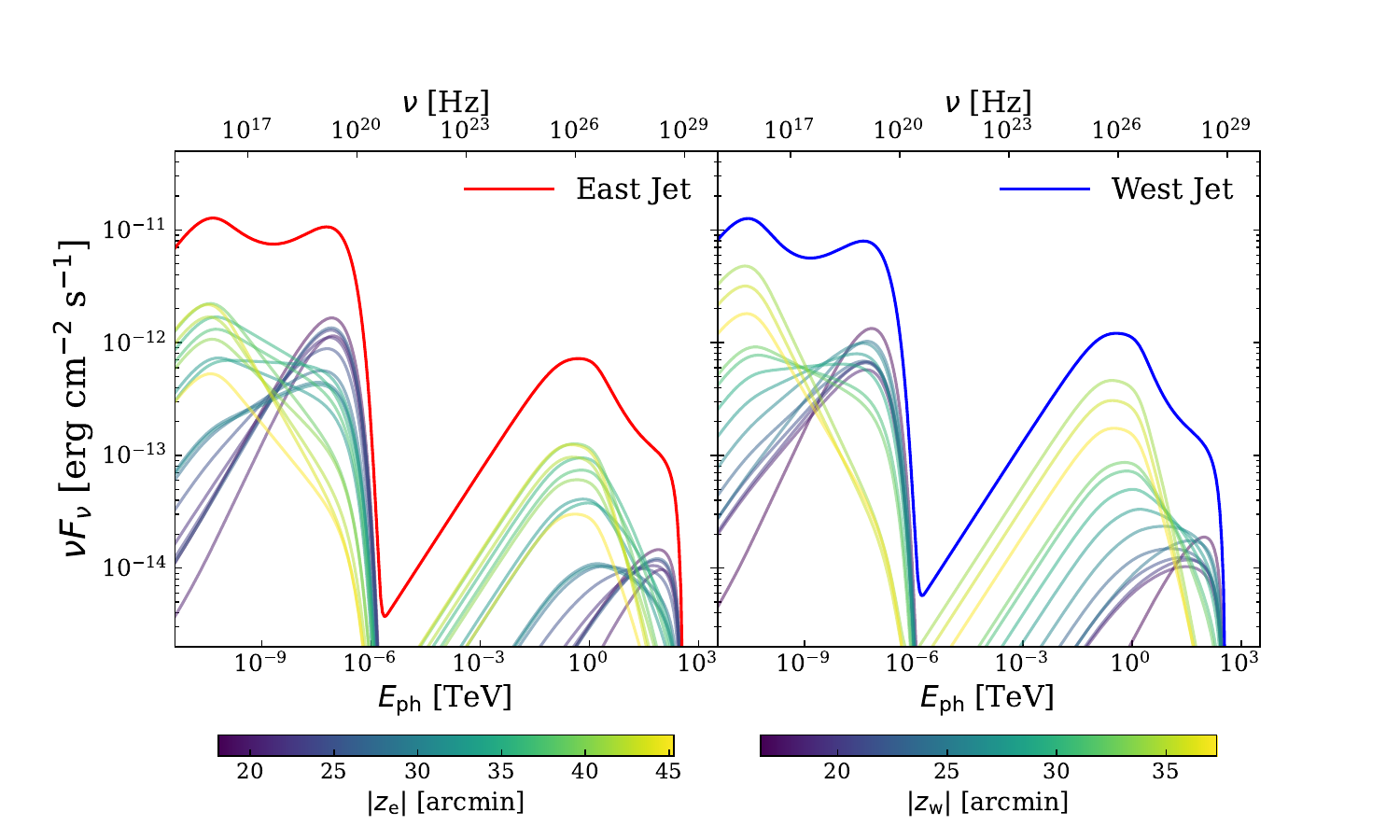}
\caption{Multiwavelength SEDs obtained from the steady-state multi-component model, showing the photon emission produced by the electron populations in the individual spatial bins. The colored curves show the contribution from the electrons in each bin, with color indicating the projected distance from the central source. The thick red and blue curves represent the total SEDs obtained by summing the contributions from all spatial bins in the eastern and western jets, respectively.}
\label{fig:sed_steady}
\end{figure*}

Figure~\ref{fig:sed_steady} illustrates  the spatially resolved SEDs given by the steady-state model shown in Section~\ref{steady-state-fitting}. The summed fluxes (the red curve in the left panel and the blue curve in the right panel) are same as those shown in the right panel of Figure~\ref{fig:steady_profile&sed}.

\bibliography{bibliography}{}

@article{stirling2004,
  author  = {Stirling, A. M. and Spencer, R. E. and Cawthorne, T. V. and Paragi, Z.},
  title   = {Polarization and Kinematic Studies of {SS 433} Indicate a Continuous and Decelerating Jet},
  journal = {Monthly Notices of the Royal Astronomical Society},
  year    = {2004},
  volume  = {354},
  number  = {4},
  pages   = {1239--1254},
  doi     = {10.1111/j.1365-2966.2004.08285.x}
}

@article{roberts2008,
  author  = {Roberts, David H. and Wardle, John F. C. and Lipnick, Scott L. and Selesnick, Philip L. and Slutsky, Simon},
  title   = {Structure and Magnetic Fields in the Precessing Jet System {SS 433}. I. Multifrequency Imaging from 1998},
  journal = {The Astrophysical Journal},
  year    = {2008},
  volume  = {676},
  number  = {1},
  pages   = {584--593},
  doi     = {10.1086/527544},
  eprint  = {0712.2005},
  archivePrefix = {arXiv},
  primaryClass  = {astro-ph}
}

@article{blundell2018,
  author  = {Blundell, Katherine M. and Laing, Robert and Lee, Steven and Richards, Anita M. S.},
  title   = {{SS433}'s Jet Trace from {ALMA} Imaging and Global Jet Watch Spectroscopy: Evidence for Post-launch Particle Acceleration},
  journal = {The Astrophysical Journal Letters},
  year    = {2018},
  volume  = {867},
  number  = {2},
  pages   = {L25},
  doi     = {10.3847/2041-8213/aae890},
  eprint  = {1811.00760},
  archivePrefix = {arXiv},
  primaryClass  = {astro-ph.HE}
}

@article{farnes2017,
  author  = {Farnes, J. S. and Gaensler, B. M. and Purcell, C. and Sun, X. H. and Haverkorn, M. and Lenc, E. and O'Sullivan, S. P. and Akahori, T.},
  title   = {Interacting Large-scale Magnetic Fields and Ionized Gas in the {W50/SS433} System},
  journal = {Monthly Notices of the Royal Astronomical Society},
  year    = {2017},
  volume  = {467},
  number  = {4},
  pages   = {4777--4801},
  doi     = {10.1093/mnras/stx338},
  eprint  = {1604.06552},
  archivePrefix = {arXiv},
  primaryClass  = {astro-ph.HE}
}

@article{sakemi2018,
  author  = {Sakemi, Haruka and Machida, Mami and Akahori, Takuya and Nakanishi, Hiroyuki and Akamatsu, Hiroki and Kurahara, Kohei and Farnes, Jamie},
  title   = {Magnetic Field Analysis of the Bow and Terminal Shock of the {SS 433} Jet},
  journal = {Publications of the Astronomical Society of Japan},
  year    = {2018},
  volume  = {70},
  number  = {2},
  pages   = {27},
  doi     = {10.1093/pasj/psy003},
  eprint  = {1802.05829},
  archivePrefix = {arXiv},
  primaryClass  = {astro-ph.HE}
}

@article{kaaret2024,
  author  = {Kaaret, Philip and Ferrazzoli, Riccardo and Silvestri, Stefano and Negro, Michela and Manfreda, Alberto and Wu, Kinwah and Costa, Enrico and Soffitta, Paolo and Safi-Harb, Samar and Poutanen, Juri and others},
  title   = {X-Ray Polarization of the Eastern Lobe of {SS 433}},
  journal = {The Astrophysical Journal Letters},
  year    = {2024},
  volume  = {961},
  number  = {1},
  pages   = {L12},
  doi     = {10.3847/2041-8213/ad103b},
  eprint  = {2311.16313},
  archivePrefix = {arXiv},
  primaryClass  = {astro-ph.HE}
}

@article{lopezmiralles2026,
  author  = {L{\'o}pez-Miralles, Jose and Perucho, Manel and Vall{\'e}s-P{\'e}rez, David and Mart{\'i}, Jos{\'e}-Mar{\'i}a and Bosch-Ramon, Valent{\'i} and Miller-Jones, James C. A. and Motta, Sara E. and Migliari, Simone and Marshall, Herman L.},
  title   = {Magnetic Field Topology and Colliding Discrete Ejecta in the Precessing Jets of {SS 433}},
  journal = {Nature Astronomy},
  year    = {2026},
  doi     = {10.1038/s41550-026-02922-6},
  note    = {Published online 14 July 2026}
}

@ARTICLE{2024Sci...383..402H,
       author = {{H.~E.~S.~S. Collaboration} and {Aharonian}, F. and {Ait Benkhali}, F. and {Aschersleben}, J. and {Ashkar}, H. and {Backes}, M. and {Barbosa Martins}, V. and {Batzofin}, R. and {Becherini}, Y. and {Berge}, D. and {Bernl{\"o}hr}, K. and {Bi}, B. and {B{\"o}ttcher}, M. and {Boisson}, C. and {Bolmont}, J. and {de Lavergne}, M. de Bony and {Borowska}, J. and {Bouyahiaoui}, M. and {Breuhaus}, M. and {Brose}, R. and {Brown}, A.~M. and {Brun}, F. and {Bruno}, B. and {Bulik}, T. and {Burger-Scheidlin}, C. and {Caroff}, S. and {Casanova}, S. and {Cecil}, R. and {Celic}, J. and {Cerruti}, M. and {Chand}, T. and {Chandra}, S. and {Chen}, A. and {Chibueze}, J. and {Chibueze}, O. and {Cotter}, G. and {Dai}, S. and {Mbarubucyeye}, J. Damascene and {Djannati-Ata{\"\i}}, A. and {Dmytriiev}, A. and {Doroshenko}, V. and {Egberts}, K. and {Einecke}, S. and {Ernenwein}, J. -P. and {Filipovic}, M. and {Fontaine}, G. and {F{\"u}{\ss}ling}, M. and {Funk}, S. and {Gabici}, S. and {Ghafourizadeh}, S. and {Giavitto}, G. and {Glawion}, D. and {Glicenstein}, J. -F. and {Grolleron}, G. and {Haerer}, L. and {Hinton}, J.~A. and {Hofmann}, W. and {Holch}, T.~L. and {Holler}, M. and {Horns}, D. and {Jamrozy}, M. and {Jankowsky}, F. and {Jardin-Blicq}, A. and {Joshi}, V. and {Jung-Richardt}, I. and {Kasai}, E. and {Katarzy{\'n}ski}, K. and {Khatoon}, R. and {Kh{\'e}lifi}, B. and {Klepser}, S. and {Klu{\'z}niak}, W. and {Komin}, Nu. and {Kosack}, K. and {Kostunin}, D. and {Kundu}, A. and {Lang}, R.~G. and {Le Stum}, S. and {Leitl}, F. and {Lemi{\`e}re}, A. and {Lenain}, J. -P. and {Leuschner}, F. and {Lohse}, T. and {Luashvili}, A. and {Lypova}, I. and {Mackey}, J. and {Malyshev}, D. and {Malyshev}, D. and {Marandon}, V. and {Marchegiani}, P. and {Marcowith}, A. and {Mart{\'\i}-Devesa}, G. and {Marx}, R. and {Mehta}, A. and {Mitchell}, A. and {Moderski}, R. and {Mohrmann}, L. and {Montanari}, A. and {Moulin}, E. and {Murach}, T. and {Nakashima}, K. and {de Naurois}, M. and {Niemiec}, J. and {Noel}, A. Priyana and {Ohm}, S. and {Olivera-Nieto}, L. and {de Ona Wilhelmi}, E. and {Ostrowski}, M. and {Panny}, S. and {Panter}, M. and {Parsons}, R.~D. and {Peron}, G. and {Prokhorov}, D.~A. and {P{\"u}hlhofer}, G. and {Punch}, M. and {Quirrenbach}, A. and {Reichherzer}, P. and {Reimer}, A. and {Reimer}, O. and {Ren}, H. and {Renaud}, M. and {Reville}, B. and {Rieger}, F. and {Rowell}, G. and {Rudak}, B. and {Ricarte}, H. Rueda and {Ruiz-Velasco}, E. and {Sahakian}, V. and {Salzmann}, H. and {Santangelo}, A. and {Sasaki}, M. and {Sch{\"a}fer}, J. and {Sch{\"u}ssler}, F. and {Schwanke}, U. and {Shapopi}, J.~N.~S. and {Sol}, H. and {Specovius}, A. and {Spencer}, S. and {Stawarz}, L. and {Steenkamp}, R. and {Steinmassl}, S. and {Steppa}, C. and {Streil}, K. and {Sushch}, I. and {Suzuki}, H. and {Takahashi}, T. and {Tanaka}, T. and {Taylor}, A.~M. and {Terrier}, R. and {Tsirou}, M. and {Tsuji}, N. and {Unbehaun}, T. and {van Eldik}, C. and {Vecchi}, M. and {Veh}, J. and {Venter}, C. and {Vink}, J. and {Wach}, T. and {Wagner}, S.~J. and {Werner}, F. and {White}, R. and {Wierzcholska}, A. and {Wong}, Yu Wun and {Zacharias}, M. and {Zargaryan}, D. and {Zdziarski}, A.~A. and {Zech}, A. and {Zouari}, S. and {{\.Z}ywucka}, N.},
        title = "{Acceleration and transport of relativistic electrons in the jets of the microquasar SS 433}",
      journal = {Science},
         year = 2024,
        month = jan,
       volume = {383},
       number = {6681},
        pages = {402-406},
          doi = {10.1126/science.adi2048},
archivePrefix = {arXiv},
       eprint = {2401.16019},
 primaryClass = {astro-ph.HE},
       adsurl = {https://ui.adsabs.harvard.edu/abs/2024Sci...383..402H}
}

@ARTICLE{Peretti2025,
       author = {{Peretti}, Enrico and {Petropoulou}, Maria and {Vasilopoulos}, Georgios and {Gabici}, Stefano},
        title = "{Particle acceleration and multi-messenger radiation from ultra-luminous X-ray sources: A new class of Galactic PeVatrons}",
      journal = {\aap},
         year = 2025,
        month = jun,
       volume = {698},
          eid = {A188},
        pages = {A188},
          doi = {10.1051/0004-6361/202452987},
archivePrefix = {arXiv},
       eprint = {2411.08762},
 primaryClass = {astro-ph.HE},
       adsurl = {https://ui.adsabs.harvard.edu/abs/2025A&A...698A.188P}
}

@ARTICLE{2020ApJ...904..188K,
       author = {{Kimura}, Shigeo S. and {Murase}, Kohta and {M{\'e}sz{\'a}ros}, Peter},
        title = "{Deciphering the Origin of the GeV-TeV Gamma-Ray Emission from SS 433}",
      journal = {\apj},
         year = 2020,
        month = dec,
       volume = {904},
       number = {2},
          eid = {188},
        pages = {188},
          doi = {10.3847/1538-4357/abbe00},
archivePrefix = {arXiv},
       eprint = {2008.04515},
 primaryClass = {astro-ph.HE},
       adsurl = {https://ui.adsabs.harvard.edu/abs/2020ApJ...904..188K}
}

@article{10.1093/nsr/nwaf496,
    author = {Cao, Zhen and Aharonian, Felix and Bai, Yun-Xiang and Bao, Yi-Wei and Bastieri, Denis and Bi, Xiao-Jun and Bi, Yu-Jiang and Bian, Wen-Yi and Bukevich, Anatoly V and Cai, Chengmiao and Cao, Wen-Yu and Cao, Zhe and Chang, Jin and Chang, Jin-Fan and Chen, Aming and Chen, En-Sheng and Chen, Guohai and Chen, Hua-Xi and Chen, Liang and Chen, Long and Chen, Ming-Jun and Chen, Ma-Li and Chen, Qi-Hui and Chen, Shi and Chen, Su-Hong and Chen, Song-Zhan and Chen, Tian-Lu and Chen, Xiao-Bin and Chen, Xuejian and Chen, Yang and Cheng, Ning and Cheng, Yao-Dong and Chu, Ming Chung and Cui, Ming-Yang and Cui, Shu-Wang and Cui, Xiao-Hong and Cui, Yi-Dong and Dai, Ben-Zhong and Dai, Hong-Liang and Dai, Zigao and Luobu, Danzeng and Diao, Yang-Xuan and Dong, Xu-Qiang and Duan, Kai-Kai and Fan, Jun-Hui and Fan, Yi-Zhong and Fang, Jun and Fang, Jian-Hua and Fang, Kun and Feng, Cun-feng and Feng, Hua and Feng, Li and Feng, Shaohui and Feng, Xiao-ting and Feng, Yi and Feng, You-liang and Gabici, Stefano and Gao, Bo and Gao, Chuan-dong and Gao, Qi and Gao, Wei and Gao, Wei-kang and Ge, Maomao and Ge, Ting-Ting and Geng, Lisi and Giacinti, Gwenael and Gong, Guanghua and Gou, Quanbu and Gu, Min-Hao and Guo, Fu-Lai and Guo, Jing and Guo, Xiao-Lei and Guo, Yi-Qing and Guo, Ying-Ying and Han, Yi-Ang and Hannuksela, Otto A and Hasan, Mariam and He, Hui-Hai and He, Hao-Ning and He, Jia-Yin and He, Xinyu and He, Yu and Hernández-Cadena, Sergio and Hou, Bo-Wen and Hou, Chao and Hou, Xian and Hu, Hong-Bo and Hu, Shi-Cong and Huang, Chen and Huang, Dai-Hui and Huang, Jiajun and Huang, Tian-Qi and Huang, Wen-Jun and Huang, Xing-Tao and Huang, Xiao-Yuan and Huang, Yong and Huang, Yi-Yun and Ji, Xiao-Lu and Jia, Huan-Yu and Jia, Kang and Jiang, Hou-Bing and Jiang, Kun and Jiang, Xiao-Wei and Jiang, Ze-Jun and Jin, Min and Kaci, Samy and Kang, Ming-Ming and Karpikov, Ivan and Khangulyan, Dmitry and Kuleshov, Denis and Kurinov, Kirill and Li, Bing-Bing and Li, Cheng and Li, Cong and Li, Dan and Li, Fei and Li, Haibo and Li, Huicai and Li, Jian and Li, Jie and Li, Kai and Li, Long and Li, Rong-Lan and Li, Si-Da and Li, Tian-Yang and Li, Wen-Lian and Li, Xiu-Rong and Li, Xin and Li, Yuan and Li, Yizhuo and Li, Zhe and Li, Zhuo and Liang, En-Wei and Liang, Yun-Feng and Lin, Su-Jie and Liu, Bing and Liu, Cheng and Liu, Dong and Liu, Dang-Bo and Liu, Hu and Liu, Hai-Dong and Liu, Jia and Liu, Jia-Li and Liu, Ji-Ren and Liu, Mao-Yuan and Liu, Ruo-Yu and Liu, Si-Ming and Liu, Wei and Liu, X and Liu, Yi and Liu, Yu and Liu, Yi-Nong and Lou, Yu-Qing and Luo, Qing and Luo, Yu and Lv, Hong-Kui and Ma, Bo-Qiang and Ma, Ling-Ling and Ma, Xin-Hua and Mao, Ji-Rong and Min, Zhen and Mitthumsiri, Warit and Mou, Guo-Bin and Mu, Hui-Jun and Neronov, Andrii and NG, Kenny Chun Yu and Ni, Ming-Yang and Nie, Lin and Ou, Le-Jian and Pattarakijwanich, Petchara and Pei, Zhi-Yuan and Qi, Jin-Can and Qi, Meng-Yao and Qin, Jia-Jun and Raza, Ali and Ren, Chong-Yang and Ruffolo, David and Sáiz, Alejandro and Semikoz, Dmitri and Shao, Lang and Shchegolev, Oleg and Shen, Yun-Zhi and Sheng, Xiang-Dong and Shi, Zhaodong and Shu, Fu-Wen and Song, Hui-Chao and Stenkin, Yuri V and Stepanov, Vladimir and Su, Yang and Sun, Dongxu and Sun, Hao and Sun, Qinning and Sun, Xiaona and Sun, Zhibin and Tabasam, Nabeel Hussain and Takata, Jumpei and Tam, Pak Hin Thomas and Tan, Hong-Bin and Tang, Qingwen and Tang, Ruiyi and Tang, Zebo and Tian, Wenwu and Tong, Chaonan and Wan, Li-Hong and Wang, Chao and Wang, Guangwei and Wang, Hongguang and Wang, Jiancheng and Wang, Ke and Wang, Kai and Wang, Liping and Wang, Lingyu and Wang, Lu-Yao and Wang, Ran and Wang, Wei and Wang, Xianggao and Wang, Xin-Jian and Wang, Xiang-Yu and Wang, Yang and Wang, Yu-Dong and Wang, Zhong-Hai and Wang, Zhong-Xiang and Wang, Zheng and Wei, Da-Ming and Wei, Jun-Jie and Wei, Yong-Jian and Wen, Tao and Weng, Shan-Shan and Wu, Chao-Yong and Wu, Han-Rong and Wu, Qing-Wen and Wu, Sha and Wu, Xue-Feng and Wu, Yu-Sheng and Xi, Shao-qiang and Xia, Jie and Xia, Jun-Ji and Xiang, Guang-man and Xiao, Di-xuan and Xiao, Gang and Xin, Yu-liang and Xing, Yi and Xiong, Ding-rong and Xiong, Zeng and Xu, Dong-lian and Xu, Reng-Feng and Xu, Ren-Xin and Xu, Wei-Li and Xue, Liang and Yan, Da-Hai and Yan, Jing-Zhi and Yan, Tian and Yang, Chao-Wen and Yang, Chu-Yuan and Yang, Feng-Fan and Yang, Li-Li and Yang, Ming-Jie and Yang, Rui-Zhi and Yang, Wen-Xin and Yang, Zihang and Yao, Zhi-Guo and Ye, Xuan-Ang and Yin, Li-Qiao and Yin, Na and You, Xiao-Hao and You, Zhi-Yong and Yu, Yan-Hong and Yuan, Qiang and Yue, Hua and Zeng, Hou-Dun and Zeng, Ting-Xuan and Zeng, Wei and Zeng, Xiangtao and Zha, Min and Zhang, Bin-Bin and Zhang, Bing Theodore and Zhang, Chao and Zhang, Feng and Zhang, Hong-Fei and Zhang, Hai-Ming and Zhang, Heng-Ying and Zhang, Jian-Li and Zhang, Li and Zhang, Peng-Fei and Zhang, Pei-Pei and Zhang, Rui and Zhang, Shao-Ru and Zhang, Shou-Shan and Zhang, Weiyan and Zhang, Xiao and Zhang, Xiao-Peng and Zhang, Yi and Zhang, Yong and Zhang, Zhi-Peng and Zhao, Jing and Zhao, Lei and Zhao, Li-Zhi and Zhao, Shi-Ping and Zhao, Xiao-Hong and Zhao, Zihao and Zheng, Fu and Zhong, Wen-Juan and Zhou, Bin and Zhou, Hao and Zhou, Jia-Neng and Zhou, Meng and Zhou, Ping and Zhou, Rong and Zhou, Xiao-Xi and Zhou, Xun-Xiu and Zhu, Ben-Yang and Zhu, Cheng-Guang and Zhu, Feng-Rong and Zhu, Hui and Zhu, Ke-Jun and Zou, Yuan-Chuan and Zuo, Xiong, The LHAASO Collaboration},
    title = {Ultrahigh-energy gamma-ray emission associated with black hole–jet systems},
    journal = {National Science Review},
    volume = {12},
    number = {12},
    pages = {nwaf496},
    year = {2025},
    month = {12},
    issn = {2095-5138},
    doi = {10.1093/nsr/nwaf496},
    url = {https://doi.org/10.1093/nsr/nwaf496},
    eprint = {https://academic.oup.com/nsr/article-pdf/12/12/nwaf496/65284677/nwaf496.pdf},
}

@ARTICLE{Popescu2017,
       author = {{Popescu}, C.~C. and {Yang}, R. and {Tuffs}, R.~J. and {Natale}, G. and {Rushton}, M. and {Aharonian}, F.},
        title = "{A radiation transfer model for the Milky Way: I. Radiation fields and application to high-energy astrophysics}",
      journal = {\mnras},
         year = 2017,
        month = sep,
       volume = {470},
       number = {3},
        pages = {2539-2558},
          doi = {10.1093/mnras/stx1282},
archivePrefix = {arXiv},
       eprint = {1705.06652},
 primaryClass = {astro-ph.GA},
       adsurl = {https://ui.adsabs.harvard.edu/abs/2017MNRAS.470.2539P}
}

@ARTICLE{2007A&A...463..611B,
       author = {{Brinkmann}, W. and {Pratt}, G.~W. and {Rohr}, S. and {Kawai}, N. and {Burwitz}, V.},
        title = "{XMM-Newton observations of the eastern jet of SS 433}",
      journal = {\aap},
         year = 2007,
        month = feb,
       volume = {463},
       number = {2},
        pages = {611-619},
          doi = {10.1051/0004-6361:20065570},
archivePrefix = {arXiv},
       eprint = {astro-ph/0610781},
 primaryClass = {astro-ph},
       adsurl = {https://ui.adsabs.harvard.edu/abs/2007A&A...463..611B}
}

@ARTICLE{2022ApJ...935..163S,
       author = {{Safi-Harb}, Samar and {Mac Intyre}, Brydyn and {Zhang}, Shuo and {Pope}, Isaac and {Zhang}, Shuhan and {Saffold}, Nathan and {Mori}, Kaya and {Gotthelf}, Eric V. and {Aharonian}, Felix and {Band}, Matthew and {Braun}, Chelsea and {Fang}, Ke and {Hailey}, Charles and {Nynka}, Melania and {Rho}, Chang D.},
        title = "{Hard X-Ray Emission from the Eastern Jet of SS 433 Powering the W50 ``Manatee'' Nebula: Evidence for Particle Reacceleration}",
      journal = {\apj},
         year = 2022,
        month = aug,
       volume = {935},
       number = {2},
          eid = {163},
        pages = {163},
          doi = {10.3847/1538-4357/ac7c05},
archivePrefix = {arXiv},
       eprint = {2207.00573},
 primaryClass = {astro-ph.HE},
       adsurl = {https://ui.adsabs.harvard.edu/abs/2022ApJ...935..163S}
}

@ARTICLE{2022PASJ...74.1143K,
       author = {{Kayama}, Kazuho and {Tanaka}, Takaaki and {Uchida}, Hiroyuki and {Tsuru}, Takeshi Go and {Sudoh}, Takahiro and {Inoue}, Yoshiyuki and {Khangulyan}, Dmitry and {Tsuji}, Naomi and {Yamamoto}, Hiroaki},
        title = "{Spatially resolved study of the SS 433/W 50 west region with Chandra: X-ray structure and spectral variation of non-thermal emission}",
      journal = {\pasj},
         year = 2022,
        month = oct,
       volume = {74},
       number = {5},
        pages = {1143-1156},
          doi = {10.1093/pasj/psac060},
archivePrefix = {arXiv},
       eprint = {2207.05924},
 primaryClass = {astro-ph.HE},
       adsurl = {https://ui.adsabs.harvard.edu/abs/2022PASJ...74.1143K}
}

@ARTICLE{2025PASJ...77..880K,
       author = {{Kayama}, Kazuho and {Tanaka}, Takaaki and {Uchida}, Hiroyuki and {Tsuru}, Takeshi Go and {Inoue}, Yoshiyuki and {Khangulyan}, Dmitry and {Tsuji}, Naomi and {Yamamoto}, Hiroaki},
        title = "{X-ray study of the propagation of non-thermal particles in microquasar SS 433/W 50 extended jets}",
      journal = {\pasj},
         year = 2025,
        month = aug,
       volume = {77},
       number = {4},
        pages = {880-889},
          doi = {10.1093/pasj/psaf059},
archivePrefix = {arXiv},
       eprint = {2505.10620},
 primaryClass = {astro-ph.HE},
       adsurl = {https://ui.adsabs.harvard.edu/abs/2025PASJ...77..880K}
}

@ARTICLE{2024ApJ...975L..28C,
       author = {{Chi}, Yi-Heng and {Huang}, Jiahui and {Zhou}, Ping and {Feng}, Hua and {Li}, Xiang-Dong and {Markoff}, Sera B. and {Safi-Harb}, Samar and {Olivera-Nieto}, Laura},
        title = "{An X-Ray Shell Reveals the Supernova Explosion for Galactic Microquasar SS 433}",
      journal = {\apjl},
         year = 2024,
        month = nov,
       volume = {975},
       number = {2},
          eid = {L28},
        pages = {L28},
          doi = {10.3847/2041-8213/ad84ed},
archivePrefix = {arXiv},
       eprint = {2410.06510},
 primaryClass = {astro-ph.HE},
       adsurl = {https://ui.adsabs.harvard.edu/abs/2024ApJ...975L..28C}
}

@ARTICLE{1998AJ....116.1842D,
       author = {{Dubner}, G.~M. and {Holdaway}, M. and {Goss}, W.~M. and {Mirabel}, I.~F.},
        title = "{A High-Resolution Radio Study of the W50-SS 433 System and the Surrounding Medium}",
      journal = {\aj},
         year = 1998,
        month = oct,
       volume = {116},
       number = {4},
        pages = {1842-1855},
          doi = {10.1086/300537},
       adsurl = {https://ui.adsabs.harvard.edu/abs/1998AJ....116.1842D}
}

@ARTICLE{2004ASPRv..12....1F,
       author = {{Fabrika}, S.},
        title = "{The jets and supercritical accretion disk in SS433}",
      journal = {\apspr},
         year = 2004,
        month = jan,
       volume = {12},
        pages = {1-152},
          doi = {10.48550/arXiv.astro-ph/0603390},
archivePrefix = {arXiv},
       eprint = {astro-ph/0603390},
 primaryClass = {astro-ph},
       adsurl = {https://ui.adsabs.harvard.edu/abs/2004ASPRv..12....1F}
}

@ARTICLE{2004ApJ...616L.159B,
       author = {{Blundell}, Katherine M. and {Bowler}, Michael G.},
        title = "{Symmetry in the Changing Jets of SS 433 and Its True Distance from Us}",
      journal = {\apjl},
         year = 2004,
        month = dec,
       volume = {616},
       number = {2},
        pages = {L159-L162},
          doi = {10.1086/426542},
archivePrefix = {arXiv},
       eprint = {astro-ph/0410456},
 primaryClass = {astro-ph},
       adsurl = {https://ui.adsabs.harvard.edu/abs/2004ApJ...616L.159B}
}

@ARTICLE{2024ApJ...976...30A,
       author = {{Alfaro}, R. and {Alvarez}, C. and {Arteaga-Vel{\'a}zquez}, J.~C. and {Avila Rojas}, D. and {Ayala Solares}, H.~A. and {Babu}, R. and {Belmont-Moreno}, E. and {Bernal}, A. and {Caballero-Mora}, K.~S. and {Capistr{\'a}n}, T. and {Carrami{\~n}ana}, A. and {Casanova}, S. and {Cotzomi}, J. and {Fuente}, E. De la and {Depaoli}, D. and {Di Lalla}, N. and {Diaz Hernandez}, R. and {Dingus}, B.~L. and {DuVernois}, M.~A. and {Engel}, K. and {Ergin}, T. and {Espinoza}, C. and {Fan}, K.~L. and {Fang}, K. and {Fraija}, N. and {Fraija}, S. and {Garc{\'\i}a-Gonz{\'a}lez}, J.~A. and {Gonzalez Mu{\~n}oz}, A. and {Gonz{\'a}lez}, M.~M. and {Goodman}, J.~A. and {Groetsch}, S. and {Harding}, J.~P. and {Hern{\'a}ndez-Cadena}, S. and {Herzog}, I. and {Huang}, D. and {Hueyotl-Zahuantitla}, F. and {H{\"u}ntemeyer}, P. and {Iriarte}, A. and {Kaufmann}, S. and {Lara}, A. and {Lee}, W.~H. and {Lee}, J. and {de Le{\'o}n}, C. and {Vargas}, H. Le{\'o}n and {Longinotti}, A.~L. and {Luis-Raya}, G. and {Malone}, K. and {Mart{\'\i}nez-Castro}, J. and {Matthews}, J.~A. and {Miranda-Romagnoli}, P. and {Montes}, J.~A. and {Moreno}, E. and {Mostaf{\'a}}, M. and {Nellen}, L. and {Nisa}, M.~U. and {Noriega-Papaqui}, R. and {P{\'e}rez Araujo}, Y. and {P{\'e}rez-P{\'e}rez}, E.~G. and {Rho}, C.~D. and {Rosa-Gonz{\'a}lez}, D. and {Ruiz-Velasco}, E. and {Salazar}, H. and {Sandoval}, A. and {Schneider}, M. and {Serna-Franco}, J. and {Smith}, A.~J. and {Son}, Y. and {Springer}, R.~W. and {Tibolla}, O. and {Tollefson}, K. and {Torres}, I. and {Torres-Escobedo}, R. and {Turner}, R. and {Ure{\~n}a-Mena}, F. and {Varela}, E. and {Villase{\~n}or}, L. and {Wang}, X. and {Wang}, Z. and {Watson}, I.~J. and {Yu}, S. and {Yun-C{\'a}rcamo}, S. and {Zhou}, H.},
        title = "{Spectral Study of Very-high-energy Gamma Rays from SS 433 with HAWC}",
      journal = {\apj},
         year = 2024,
        month = nov,
       volume = {976},
       number = {1},
          eid = {30},
        pages = {30},
          doi = {10.3847/1538-4357/ad7e1b},
archivePrefix = {arXiv},
       eprint = {2410.21796},
 primaryClass = {astro-ph.HE},
       adsurl = {https://ui.adsabs.harvard.edu/abs/2024ApJ...976...30A}
}

@ARTICLE{2020NatAs...4.1177L,
       author = {{Li}, Jian and {Torres}, Diego F. and {Liu}, Ruo-Yu and {Kerr}, Matthew and {de O{\~n}a Wilhelmi}, Emma and {Su}, Yang},
        title = "{Gamma-ray heartbeat powered by the microquasar SS 433}",
      journal = {Nature Astronomy},
         year = 2020,
        month = jan,
       volume = {4},
        pages = {1177-1184},
          doi = {10.1038/s41550-020-1164-6},
archivePrefix = {arXiv},
       eprint = {2008.10523},
 primaryClass = {astro-ph.HE},
       adsurl = {https://ui.adsabs.harvard.edu/abs/2020NatAs...4.1177L}
}

@ARTICLE{1983ApJ...273..688W,
       author = {{Watson}, M.~G. and {Willingale}, R. and {Grindlay}, J.~E. and {Seward}, F.~D.},
        title = "{The X-ray lobes of SS 433.}",
      journal = {\apj},
         year = 1983,
        month = oct,
       volume = {273},
        pages = {688-696},
          doi = {10.1086/161403},
       adsurl = {https://ui.adsabs.harvard.edu/abs/1983ApJ...273..688W}
}

@ARTICLE{1994PASJ...46L.109Y,
       author = {{Yamauchi}, Shigeo and {Kawai}, Nobuyuki and {Aoki}, Takashi},
        title = "{A Non-Thermal X-Ray Spectrum from the Supernova Remnant W 50}",
      journal = {\pasj},
         year = 1994,
        month = jun,
       volume = {46},
       number = {3},
        pages = {L109-L113},
          doi = {10.1093/pasj/46.3.109},
       adsurl = {https://ui.adsabs.harvard.edu/abs/1994PASJ...46L.109Y}
}

@ARTICLE{1996A&A...312..306B,
       author = {{Brinkmann}, W. and {Aschenbach}, B. and {Kawai}, N.},
        title = "{ROSAT observations of the W 50/SS 433 system.}",
      journal = {\aap},
         year = 1996,
        month = aug,
       volume = {312},
        pages = {306-316},
       adsurl = {https://ui.adsabs.harvard.edu/abs/1996A&A...312..306B}
}

@ARTICLE{1997ApJ...483..868S,
       author = {{Safi-Harb}, Samar and {{\"O}gelman}, Hakki},
        title = "{ROSAT and ASCA Observations of W50 Associated with the Peculiar Source SS 433}",
      journal = {\apj},
         year = 1997,
        month = jul,
       volume = {483},
       number = {2},
        pages = {868-881},
          doi = {10.1086/304274},
       adsurl = {https://ui.adsabs.harvard.edu/abs/1997ApJ...483..868S}
}

@ARTICLE{2020ApJ...889..146S,
       author = {{Sudoh}, Takahiro and {Inoue}, Yoshiyuki and {Khangulyan}, Dmitry},
        title = "{Multiwavelength Emission from Galactic Jets: The Case of the Microquasar SS433}",
      journal = {\apj},
         year = 2020,
        month = feb,
       volume = {889},
       number = {2},
          eid = {146},
        pages = {146},
          doi = {10.3847/1538-4357/ab6442},
archivePrefix = {arXiv},
       eprint = {1911.00013},
 primaryClass = {astro-ph.HE},
       adsurl = {https://ui.adsabs.harvard.edu/abs/2020ApJ...889..146S}
}

@ARTICLE{2007MNRAS.381..881L,
       author = {{Lockman}, Felix J. and {Blundell}, Katherine M. and {Goss}, W.~M.},
        title = "{The distance to SS433/W50 and its interaction with the interstellar medium}",
      journal = {\mnras},
         year = 2007,
        month = nov,
       volume = {381},
       number = {3},
        pages = {881-893},
          doi = {10.1111/j.1365-2966.2007.12170.x},
archivePrefix = {arXiv},
       eprint = {0707.0506},
 primaryClass = {astro-ph},
       adsurl = {https://ui.adsabs.harvard.edu/abs/2007MNRAS.381..881L}
}

@ARTICLE{2001ApJ...548..820B,
       author = {{Borkowski}, Kazimierz J. and {Lyerly}, William J. and {Reynolds}, Stephen P.},
        title = "{Supernova Remnants in the Sedov Expansion Phase: Thermal X-Ray Emission}",
      journal = {\apj},
         year = 2001,
        month = feb,
       volume = {548},
       number = {2},
        pages = {820-835},
          doi = {10.1086/319011},
archivePrefix = {arXiv},
       eprint = {astro-ph/0008066},
 primaryClass = {astro-ph},
       adsurl = {https://ui.adsabs.harvard.edu/abs/2001ApJ...548..820B}
}

@ARTICLE{2024Natur.634..557A,
       author = {{Alfaro}, R. and {Alvarez}, C. and {Arteaga-Vel{\'a}zquez}, J.~C. and {Avila Rojas}, D. and {Ayala Solares}, H.~A. and {Babu}, R. and {Belmont-Moreno}, E. and {Caballero-Mora}, K.~S. and {Capistr{\'a}n}, T. and {Carrami{\~n}ana}, A. and {Casanova}, S. and {Cotti}, U. and {Cotzomi}, J. and {Couti{\~n}o de Le{\'o}n}, S. and {De la Fuente}, E. and {Depaoli}, D. and {Di Lalla}, N. and {Diaz Hernandez}, R. and {Dingus}, B.~L. and {DuVernois}, M.~A. and {Durocher}, M. and {D{\'\i}az-V{\'e}lez}, J.~C. and {Engel}, K. and {Espinoza}, C. and {Fan}, K.~L. and {Fang}, K. and {Fraija}, N. and {Fraija}, S. and {Garc{\'\i}a-Gonz{\'a}lez}, J.~A. and {Garfias}, F. and {Gonzalez Mu{\~n}oz}, A. and {Gonz{\'a}lez}, M.~M. and {Goodman}, J.~A. and {Groetsch}, S. and {Harding}, J.~P. and {Herzog}, I. and {Hinton}, J. and {Huang}, D. and {Hueyotl-Zahuantitla}, F. and {H{\"u}ntemeyer}, P. and {Iriarte}, A. and {Joshi}, V. and {Kaufmann}, S. and {Kieda}, D. and {de Le{\'o}n}, C. and {Lee}, J. and {Le{\'o}n Vargas}, H. and {Linnemann}, J.~T. and {Longinotti}, A.~L. and {Luis-Raya}, G. and {Malone}, K. and {Martinez}, O. and {Mart{\'\i}nez-Castro}, J. and {Matthews}, J.~A. and {Miranda-Romagnoli}, P. and {Morales-Soto}, J.~A. and {Moreno}, E. and {Mostaf{\'a}}, M. and {Nayerhoda}, A. and {Nellen}, L. and {Newbold}, M. and {Nisa}, M.~U. and {Noriega-Papaqui}, R. and {Olivera-Nieto}, L. and {Omodei}, N. and {Osorio}, M. and {P{\'e}rez Araujo}, Y. and {P{\'e}rez-P{\'e}rez}, E.~G. and {Rho}, C.~D. and {Rosa-Gonz{\'a}lez}, D. and {Ruiz-Velasco}, E. and {Salazar}, H. and {Salazar-Gallegos}, D. and {Sandoval}, A. and {Schneider}, M. and {Serna-Franco}, J. and {Smith}, A.~J. and {Son}, Y. and {Springer}, R.~W. and {Tibolla}, O. and {Tollefson}, K. and {Torres}, I. and {Torres-Escobedo}, R. and {Turner}, R. and {Ure{\~n}a-Mena}, F. and {Varela}, E. and {Villase{\~n}or}, L. and {Wang}, X. and {Watson}, I.~J. and {Willox}, E. and {Yun-C{\'a}rcamo}, S. and {Zhou}, H.},
        title = "{Ultra-high-energy gamma-ray bubble around microquasar V4641 Sgr}",
      journal = {\nat},
         year = 2024,
        month = oct,
       volume = {634},
       number = {8034},
        pages = {557-560},
          doi = {10.1038/s41586-024-07995-9},
archivePrefix = {arXiv},
       eprint = {2410.16117},
 primaryClass = {astro-ph.HE},
       adsurl = {https://ui.adsabs.harvard.edu/abs/2024Natur.634..557A}
}

@ARTICLE{2006A&A...453..387P,
       author = {{Parizot}, E. and {Marcowith}, A. and {Ballet}, J. and {Gallant}, Y.~A.},
        title = "{Observational constraints on energetic particle diffusion in young supernovae remnants: amplified magnetic field and maximum energy}",
      journal = {\aap},
         year = 2006,
        month = jul,
       volume = {453},
       number = {2},
        pages = {387-395},
          doi = {10.1051/0004-6361:20064985},
archivePrefix = {arXiv},
       eprint = {astro-ph/0603723},
 primaryClass = {astro-ph},
       adsurl = {https://ui.adsabs.harvard.edu/abs/2006A&A...453..387P}
}

@ARTICLE{1978ApJ...221L..29B,
       author = {{Blandford}, R.~D. and {Ostriker}, J.~P.},
        title = "{Particle acceleration by astrophysical shocks.}",
      journal = {\apjl},
         year = 1978,
        month = apr,
       volume = {221},
        pages = {L29-L32},
          doi = {10.1086/182658},
       adsurl = {https://ui.adsabs.harvard.edu/abs/1978ApJ...221L..29B}
}

@ARTICLE{1987PhR...154....1B,
       author = {{Blandford}, Roger and {Eichler}, David},
        title = "{Particle acceleration at astrophysical shocks: A theory of cosmic ray origin}",
      journal = {\physrep},
         year = 1987,
        month = oct,
       volume = {154},
       number = {1},
        pages = {1-75},
          doi = {10.1016/0370-1573(87)90134-7},
       adsurl = {https://ui.adsabs.harvard.edu/abs/1987PhR...154....1B}
}

@ARTICLE{2001JPhG...27.1589K,
       author = {{Kirk}, J.~G. and {Dendy}, R.~O.},
        title = "{Shock acceleration of cosmic rays - a critical review}",
      journal = {Journal of Physics G Nuclear Physics},
         year = 2001,
        month = jul,
       volume = {27},
       number = {7},
        pages = {1589-1595},
          doi = {10.1088/0954-3899/27/7/316},
archivePrefix = {arXiv},
       eprint = {astro-ph/0101175},
 primaryClass = {astro-ph},
       adsurl = {https://ui.adsabs.harvard.edu/abs/2001JPhG...27.1589K}
}

@ARTICLE{2007Ap&SS.309..119R,
       author = {{Rieger}, Frank M. and {Bosch-Ramon}, Valent{\'\i} and {Duffy}, Peter},
        title = "{Fermi acceleration in astrophysical jets}",
      journal = {\apss},
         year = 2007,
        month = jun,
       volume = {309},
       number = {1-4},
        pages = {119-125},
          doi = {10.1007/s10509-007-9466-z},
archivePrefix = {arXiv},
       eprint = {astro-ph/0610141},
 primaryClass = {astro-ph},
       adsurl = {https://ui.adsabs.harvard.edu/abs/2007Ap&SS.309..119R}
}

@ARTICLE{2015ApJ...807L...8B,
       author = {{Bordas}, P. and {Yang}, R. and {Kafexhiu}, E. and {Aharonian}, F.},
        title = "{Detection of Persistent Gamma-Ray Emission Toward SS433/W50}",
      journal = {\apjl},
         year = 2015,
        month = jul,
       volume = {807},
       number = {1},
          eid = {L8},
        pages = {L8},
          doi = {10.1088/2041-8205/807/1/L8},
archivePrefix = {arXiv},
       eprint = {1411.7413},
 primaryClass = {astro-ph.HE},
       adsurl = {https://ui.adsabs.harvard.edu/abs/2015ApJ...807L...8B}
}

@ARTICLE{2018Natur.564E..38A,
       author = {{Abeysekara}, A.~U. and {Albert}, A. and {Alfaro}, R. and {Alvarez}, C. and {{\'A}lvarez}, J.~D. and {Arceo}, R. and {Arteaga-Vel{\'a}zquez}, J.~C. and {Avila Rojas}, D. and {Ayala Solares}, H.~A. and {Belmont-Moreno}, E. and {BenZvi}, S.~Y. and {Brisbois}, C. and {Caballero-Mora}, K.~S. and {Capistr{\'a}n}, T. and {Carrami{\~n}ana}, A. and {Casanova}, S. and {Castillo}, M. and {Cotti}, U. and {Cotzomi}, J. and {Couti{\~n}o de Le{\'o}n}, S. and {De Le{\'o}n}, C. and {De la Fuente}, E. and {D{\'\i}az-V{\'e}lez}, J.~C. and {Dichiara}, S. and {Dingus}, B.~L. and {DuVernois}, M.~A. and {Ellsworth}, R.~W. and {Engel}, K. and {Espinoza}, C. and {Fang}, K. and {Fleischhack}, H. and {Fraija}, N. and {Galv{\'a}n-G{\'a}mez}, A. and {Garc{\'\i}a-Gonz{\'a}lez}, J.~A. and {Garfias}, F. and {Gonz{\'a}lez-Mu{\~n}oz}, A. and {Gonz{\'a}lez}, M.~M. and {Goodman}, J.~A. and {Hampel-Arias}, Z. and {Harding}, J.~P. and {Hernandez}, S. and {Hinton}, J. and {Hona}, B. and {Hueyotl-Zahuantitla}, F. and {Hui}, C.~M. and {H{\"u}ntemeyer}, P. and {Iriarte}, A. and {Jardin-Blicq}, A. and {Joshi}, V. and {Kaufmann}, S. and {Kar}, P. and {Kunde}, G.~J. and {Lauer}, R.~J. and {Lee}, W.~H. and {Le{\'o}n Vargas}, H. and {Li}, H. and {Linnemann}, J.~T. and {Longinotti}, A.~L. and {Luis-Raya}, G. and {L{\'o}pez-Coto}, R. and {Malone}, K. and {Marinelli}, S.~S. and {Martinez}, O. and {Martinez-Castellanos}, I. and {Mart{\'\i}nez-Castro}, J. and {Matthews}, J.~A. and {Miranda-Romagnoli}, P. and {Moreno}, E. and {Mostaf{\'a}}, M. and {Nayerhoda}, A. and {Nellen}, L. and {Newbold}, M. and {Nisa}, M.~U. and {Noriega-Papaqui}, R. and {Pretz}, J. and {P{\'e}rez-P{\'e}rez}, E.~G. and {Ren}, Z. and {Rho}, C.~D. and {Rivi{\`e}re}, C. and {Rosa-Gonz{\'a}lez}, D. and {Rosenberg}, M. and {Ruiz-Velasco}, E. and {Salesa Greus}, F. and {Sandoval}, A. and {Schneider}, M. and {Schoorlemmer}, H. and {Seglar Arroyo}, M. and {Sinnis}, G. and {Smith}, A.~J. and {Springer}, R.~W. and {Surajbali}, P. and {Taboada}, I. and {Tibolla}, O. and {Tollefson}, K. and {Torres}, I. and {Vianello}, G. and {Villase{\~n}or}, L. and {Weisgarber}, T. and {Werner}, F. and {Westerhoff}, S. and {Wood}, J. and {Yapici}, T. and {Yodh}, G. and {Zepeda}, A. and {Zhang}, H. and {Zhou}, H.},
        title = "{Publisher Correction: Very-high-energy particle acceleration powered by the jets of the microquasar SS 433}",
      journal = {\nat},
         year = 2018,
        month = nov,
       volume = {564},
       number = {7736},
        pages = {E38-E38},
          doi = {10.1038/s41586-018-0688-8},
       adsurl = {https://ui.adsabs.harvard.edu/abs/2018Natur.564E..38A}
}

@ARTICLE{2019A&A...626A.113S,
       author = {{Sun}, Xiao-Na and {Yang}, Rui-Zhi and {Liu}, Bing and {Xi}, Shao-Qiang and {Wang}, Xiang-Yu},
        title = "{Tentative evidence of spatially extended GeV emission from SS433/W50}",
      journal = {\aap},
         year = 2019,
        month = jun,
       volume = {626},
          eid = {A113},
        pages = {A113},
          doi = {10.1051/0004-6361/201935621},
archivePrefix = {arXiv},
       eprint = {1904.05127},
 primaryClass = {astro-ph.HE},
       adsurl = {https://ui.adsabs.harvard.edu/abs/2019A&A...626A.113S}
}

@ARTICLE{2021MNRAS.507L..19C,
       author = {{Cherepashchuk}, A.~M. and {Belinski}, A.~A. and {Dodin}, A.~V. and {Postnov}, K.~A.},
        title = "{Discovery of orbital eccentricity and evidence for orbital period increase of SS433}",
      journal = {\mnras},
         year = 2021,
        month = oct,
       volume = {507},
       number = {1},
        pages = {L19-L23},
          doi = {10.1093/mnrasl/slab083},
archivePrefix = {arXiv},
       eprint = {2107.09005},
 primaryClass = {astro-ph.SR},
       adsurl = {https://ui.adsabs.harvard.edu/abs/2021MNRAS.507L..19C}
}

@ARTICLE{2018ApJ...863..103S,
       author = {{Su}, Yang and {Zhou}, Xin and {Yang}, Ji and {Chen}, Yang and {Chen}, Xuepeng and {Zhang}, Shaobo},
        title = "{The Large-scale Interstellar Medium of SS 433/W50 Revisited}",
      journal = {\apj},
         year = 2018,
        month = aug,
       volume = {863},
       number = {1},
          eid = {103},
        pages = {103},
          doi = {10.3847/1538-4357/aad04e},
archivePrefix = {arXiv},
       eprint = {1807.03737},
 primaryClass = {astro-ph.GA},
       adsurl = {https://ui.adsabs.harvard.edu/abs/2018ApJ...863..103S}
}

@article{doi:10.1142/S0218271810016646,
author = {Bordas, POL and BOSCH-RAMON, VALENT\'{I} and PAREDES, JOSEP MARIA},
title = {GAMMA-RAYS FROM SS 433 AND ITS INTERACTION WITH THE W50 NEBULA},
journal = {International Journal of Modern Physics D},
volume = {19},
number = {06},
pages = {749-755},
year = {2010},
doi = {10.1142/S0218271810016646},

URL = { 
    
        https://doi.org/10.1142/S0218271810016646
    
    

},
eprint = { 
    
        https://doi.org/10.1142/S0218271810016646
    
    

}
}

@INPROCEEDINGS{1996ASPC..101...17A,
       author = {{Arnaud}, K.~A.},
        title = "{XSPEC: The First Ten Years}",
    booktitle = {Astronomical Data Analysis Software and Systems V},
         year = 1996,
       editor = {{Jacoby}, George H. and {Barnes}, Jeannette},
       series = {Astronomical Society of the Pacific Conference Series},
       volume = {101},
        month = jan,
        pages = {17},
       adsurl = {https://ui.adsabs.harvard.edu/abs/1996ASPC..101...17A}
}

@ARTICLE{2018A&A...619L...4B,
       author = {{Bowler}, M.~G.},
        title = "{SS 433: Two robust determinations fix the mass ratio}",
      journal = {\aap},
         year = 2018,
        month = nov,
       volume = {619},
          eid = {L4},
        pages = {L4},
          doi = {10.1051/0004-6361/201834121},
       adsurl = {https://ui.adsabs.harvard.edu/abs/2018A&A...619L...4B}
}

@ARTICLE{2024ApJS..271...25C,
       author = {{Cao}, Zhen and {Aharonian}, F. and {An}, Q. and {Axikegu} and {Bai}, Y.~X. and {Bao}, Y.~W. and {Bastieri}, D. and {Bi}, X.~J. and {Bi}, Y.~J. and {Cai}, J.~T. and {Cao}, Q. and {Cao}, W.~Y. and {Cao}, Zhe and {Chang}, J. and {Chang}, J.~F. and {Chen}, A.~M. and {Chen}, E.~S. and {Chen}, Liang and {Chen}, Lin and {Chen}, Long and {Chen}, M.~J. and {Chen}, M.~L. and {Chen}, Q.~H. and {Chen}, S.~H. and {Chen}, S.~Z. and {Chen}, T.~L. and {Chen}, Y. and {Cheng}, N. and {Cheng}, Y.~D. and {Cui}, M.~Y. and {Cui}, S.~W. and {Cui}, X.~H. and {Cui}, Y.~D. and {Dai}, B.~Z. and {Dai}, H.~L. and {Dai}, Z.~G. and {Danzengluobu} and {Della Volpe}, D. and {Dong}, X.~Q. and {Duan}, K.~K. and {Fan}, J.~H. and {Fan}, Y.~Z. and {Fang}, J. and {Fang}, K. and {Feng}, C.~F. and {Feng}, L. and {Feng}, S.~H. and {Feng}, X.~T. and {Feng}, Y.~L. and {Gabici}, S. and {Gao}, B. and {Gao}, C.~D. and {Gao}, L.~Q. and {Gao}, Q. and {Gao}, W. and {Gao}, W.~K. and {Ge}, M.~M. and {Geng}, L.~S. and {Giacinti}, G. and {Gong}, G.~H. and {Gou}, Q.~B. and {Gu}, M.~H. and {Guo}, F.~L. and {Guo}, X.~L. and {Guo}, Y.~Q. and {Guo}, Y.~Y. and {Han}, Y.~A. and {He}, H.~H. and {He}, H.~N. and {He}, J.~Y. and {He}, X.~B. and {He}, Y. and {Heller}, M. and {Hor}, Y.~K. and {Hou}, B.~W. and {Hou}, C. and {Hou}, X. and {Hu}, H.~B. and {Hu}, Q. and {Hu}, S.~C. and {Huang}, D.~H. and {Huang}, T.~Q. and {Huang}, W.~J. and {Huang}, X.~T. and {Huang}, X.~Y. and {Huang}, Y. and {Huang}, Z.~C. and {Ji}, X.~L. and {Jia}, H.~Y. and {Jia}, K. and {Jiang}, K. and {Jiang}, X.~W. and {Jiang}, Z.~J. and {Jin}, M. and {Kang}, M.~M. and {Ke}, T. and {Kuleshov}, D. and {Kurinov}, K. and {Li}, B.~B. and {Li}, Cheng and {Li}, Cong and {Li}, D. and {Li}, F. and {Li}, H.~B. and {Li}, H.~C. and {Li}, H.~Y. and {Li}, J. and {Li}, Jian and {Li}, Jie and {Li}, K. and {Li}, W.~L. and {Li}, W.~L. and {Li}, X.~R. and {Li}, Xin and {Li}, Y.~Z. and {Li}, Zhe and {Li}, Zhuo and {Liang}, E.~W. and {Liang}, Y.~F. and {Lin}, S.~J. and {Liu}, B. and {Liu}, C. and {Liu}, D. and {Liu}, H. and {Liu}, H.~D. and {Liu}, J. and {Liu}, J.~L. and {Liu}, J.~Y. and {Liu}, M.~Y. and {Liu}, R.~Y. and {Liu}, S.~M. and {Liu}, W. and {Liu}, Y. and {Liu}, Y.~N. and {Lu}, R. and {Luo}, Q. and {Lv}, H.~K. and {Ma}, B.~Q. and {Ma}, L.~L. and {Ma}, X.~H. and {Mao}, J.~R. and {Min}, Z. and {Mitthumsiri}, W. and {Mu}, H.~J. and {Nan}, Y.~C. and {Neronov}, A. and {Ou}, Z.~W. and {Pang}, B.~Y. and {Pattarakijwanich}, P. and {Pei}, Z.~Y. and {Qi}, M.~Y. and {Qi}, Y.~Q. and {Qiao}, B.~Q. and {Qin}, J.~J. and {Ruffolo}, D. and {S{\'a}iz}, A. and {Semikoz}, D. and {Shao}, C.~Y. and {Shao}, L. and {Shchegolev}, O. and {Sheng}, X.~D. and {Shu}, F.~W. and {Song}, H.~C. and {Stenkin}, Yu. V. and {Stepanov}, V. and {Su}, Y. and {Sun}, Q.~N. and {Sun}, X.~N. and {Sun}, Z.~B. and {Tam}, P.~H.~T. and {Tang}, Q.~W. and {Tang}, Z.~B. and {Tian}, W.~W. and {Wang}, C. and {Wang}, C.~B. and {Wang}, G.~W. and {Wang}, H.~G. and {Wang}, H.~H. and {Wang}, J.~C. and {Wang}, K. and {Wang}, L.~P. and {Wang}, L.~Y. and {Wang}, P.~H. and {Wang}, R. and {Wang}, W. and {Wang}, X.~G. and {Wang}, X.~Y. and {Wang}, Y. and {Wang}, Y.~D. and {Wang}, Y.~J. and {Wang}, Z.~H. and {Wang}, Z.~X. and {Wang}, Zhen and {Wang}, Zheng and {Wei}, D.~M. and {Wei}, J.~J. and {Wei}, Y.~J. and {Wen}, T. and {Wu}, C.~Y. and {Wu}, H.~R.},
        title = "{The First LHAASO Catalog of Gamma-Ray Sources}",
      journal = {\apjs},
         year = 2024,
        month = mar,
       volume = {271},
       number = {1},
          eid = {25},
        pages = {25},
          doi = {10.3847/1538-4365/acfd29},
archivePrefix = {arXiv},
       eprint = {2305.17030},
 primaryClass = {astro-ph.HE},
       adsurl = {https://ui.adsabs.harvard.edu/abs/2024ApJS..271...25C}
}

@ARTICLE{1997MNRAS.285..449C,
       author = {{Chen}, L.-W. and {Fabian}, A.~C. and {Gendreau}, K.~C.},
        title = "{ASCA and ROSAT observations of the QSF3 field: the X-ray background in the 0.1-7 keV band}",
      journal = {\mnras},
         year = 1997,
        month = mar,
       volume = {285},
       number = {3},
        pages = {449-471},
          doi = {10.1093/mnras/285.3.449},
       adsurl = {https://ui.adsabs.harvard.edu/abs/1997MNRAS.285..449C}
}

@ARTICLE{2019MNRAS.485.2638C,
       author = {{Cherepashchuk}, A.~M. and {Postnov}, K.~A. and {Belinski}, A.~A.},
        title = "{Mass ratio in SS433 revisited}",
      journal = {\mnras},
         year = 2019,
        month = may,
       volume = {485},
       number = {2},
        pages = {2638-2641},
          doi = {10.1093/mnras/stz610},
archivePrefix = {arXiv},
       eprint = {1902.11137},
 primaryClass = {astro-ph.HE},
       adsurl = {https://ui.adsabs.harvard.edu/abs/2019MNRAS.485.2638C}
}

@ARTICLE{2020ApJ...889L...5F,
       author = {{Fang}, Ke and {Charles}, Eric and {Blandford}, Roger D.},
        title = "{GeV-TeV Counterparts of SS 433/W50 from Fermi-LAT and HAWC Observations}",
      journal = {\apjl},
         year = 2020,
        month = jan,
       volume = {889},
       number = {1},
          eid = {L5},
        pages = {L5},
          doi = {10.3847/2041-8213/ab62b8},
archivePrefix = {arXiv},
       eprint = {2001.03599},
 primaryClass = {astro-ph.HE},
       adsurl = {https://ui.adsabs.harvard.edu/abs/2020ApJ...889L...5F}
}

@INPROCEEDINGS{2004ASPC..314..759G,
       author = {{Gabriel}, C. and {Denby}, M. and {Fyfe}, D.~J. and {Hoar}, J. and {Ibarra}, A. and {Ojero}, E. and {Osborne}, J. and {Saxton}, R.~D. and {Lammers}, U. and {Vacanti}, G.},
        title = "{The XMM-Newton SAS - Distributed Development and Maintenance of a Large Science Analysis System: A Critical Analysis}",
    booktitle = {Astronomical Data Analysis Software and Systems (ADASS) XIII},
         year = 2004,
       editor = {{Ochsenbein}, Francois and {Allen}, Mark G. and {Egret}, Daniel},
       series = {Astronomical Society of the Pacific Conference Series},
       volume = {314},
        month = jul,
        pages = {759},
       adsurl = {https://ui.adsabs.harvard.edu/abs/2004ASPC..314..759G}
}

@ARTICLE{2002ApJ...578L..67G,
       author = {{Gies}, D.~R. and {Huang}, W. and {McSwain}, M.~V.},
        title = "{The Spectrum of the Mass Donor Star in SS 433}",
      journal = {\apjl},
         year = 2002,
        month = oct,
       volume = {578},
       number = {1},
        pages = {L67-L70},
          doi = {10.1086/344436},
archivePrefix = {arXiv},
       eprint = {astro-ph/0208044},
 primaryClass = {astro-ph},
       adsurl = {https://ui.adsabs.harvard.edu/abs/2002ApJ...578L..67G}
}

@ARTICLE{2001A&A...365L...1J,
       author = {{Jansen}, F. and {Lumb}, D. and {Altieri}, B. and {Clavel}, J. and {Ehle}, M. and {Erd}, C. and {Gabriel}, C. and {Guainazzi}, M. and {Gondoin}, P. and {Much}, R. and {Munoz}, R. and {Santos}, M. and {Schartel}, N. and {Texier}, D. and {Vacanti}, G.},
        title = "{XMM-Newton observatory. I. The spacecraft and operations}",
      journal = {\aap},
         year = 2001,
        month = jan,
       volume = {365},
        pages = {L1-L6},
          doi = {10.1051/0004-6361:20000036},
       adsurl = {https://ui.adsabs.harvard.edu/abs/2001A&A...365L...1J}
}

@ARTICLE{2008A&A...478..575K,
       author = {{Kuntz}, K.~D. and {Snowden}, S.~L.},
        title = "{The EPIC-MOS particle-induced background spectra}",
      journal = {\aap},
         year = 2008,
        month = feb,
       volume = {478},
       number = {2},
        pages = {575-596},
          doi = {10.1051/0004-6361:20077912},
       adsurl = {https://ui.adsabs.harvard.edu/abs/2008A&A...478..575K}
}

@ARTICLE{2004ApJ...610.1182S,
       author = {{Snowden}, S.~L. and {Collier}, M.~R. and {Kuntz}, K.~D.},
        title = "{XMM-Newton Observation of Solar Wind Charge Exchange Emission}",
      journal = {\apj},
         year = 2004,
        month = aug,
       volume = {610},
       number = {2},
        pages = {1182-1190},
          doi = {10.1086/421841},
archivePrefix = {arXiv},
       eprint = {astro-ph/0404354},
 primaryClass = {astro-ph},
       adsurl = {https://ui.adsabs.harvard.edu/abs/2004ApJ...610.1182S}
}

@ARTICLE{2001A&A...365L..18S,
       author = {{Str{\"u}der}, L. and {Briel}, U. and {Dennerl}, K. and {Hartmann}, R. and {Kendziorra}, E. and {Meidinger}, N. and {Pfeffermann}, E. and {Reppin}, C. and {Aschenbach}, B. and {Bornemann}, W. and {Br{\"a}uninger}, H. and {Burkert}, W. and {Elender}, M. and {Freyberg}, M. and {Haberl}, F. and {Hartner}, G. and {Heuschmann}, F. and {Hippmann}, H. and {Kastelic}, E. and {Kemmer}, S. and {Kettenring}, G. and {Kink}, W. and {Krause}, N. and {M{\"u}ller}, S. and {Oppitz}, A. and {Pietsch}, W. and {Popp}, M. and {Predehl}, P. and {Read}, A. and {Stephan}, K.~H. and {St{\"o}tter}, D. and {Tr{\"u}mper}, J. and {Holl}, P. and {Kemmer}, J. and {Soltau}, H. and {St{\"o}tter}, R. and {Weber}, U. and {Weichert}, U. and {von Zanthier}, C. and {Carathanassis}, D. and {Lutz}, G. and {Richter}, R.~H. and {Solc}, P. and {B{\"o}ttcher}, H. and {Kuster}, M. and {Staubert}, R. and {Abbey}, A. and {Holland}, A. and {Turner}, M. and {Balasini}, M. and {Bignami}, G.~F. and {La Palombara}, N. and {Villa}, G. and {Buttler}, W. and {Gianini}, F. and {Lain{\'e}}, R. and {Lumb}, D. and {Dhez}, P.},
        title = "{The European Photon Imaging Camera on XMM-Newton: The pn-CCD camera}",
      journal = {\aap},
         year = 2001,
        month = jan,
       volume = {365},
        pages = {L18-L26},
          doi = {10.1051/0004-6361:20000066},
       adsurl = {https://ui.adsabs.harvard.edu/abs/2001A&A...365L..18S}
}

@ARTICLE{2001A&A...365L..27T,
       author = {{Turner}, M.~J.~L. and {Abbey}, A. and {Arnaud}, M. and {Balasini}, M. and {Barbera}, M. and {Belsole}, E. and {Bennie}, P.~J. and {Bernard}, J.~P. and {Bignami}, G.~F. and {Boer}, M. and {Briel}, U. and {Butler}, I. and {Cara}, C. and {Chabaud}, C. and {Cole}, R. and {Collura}, A. and {Conte}, M. and {Cros}, A. and {Denby}, M. and {Dhez}, P. and {Di Coco}, G. and {Dowson}, J. and {Ferrando}, P. and {Ghizzardi}, S. and {Gianotti}, F. and {Goodall}, C.~V. and {Gretton}, L. and {Griffiths}, R.~G. and {Hainaut}, O. and {Hochedez}, J.~F. and {Holland}, A.~D. and {Jourdain}, E. and {Kendziorra}, E. and {Lagostina}, A. and {Laine}, R. and {La Palombara}, N. and {Lortholary}, M. and {Lumb}, D. and {Marty}, P. and {Molendi}, S. and {Pigot}, C. and {Poindron}, E. and {Pounds}, K.~A. and {Reeves}, J.~N. and {Reppin}, C. and {Rothenflug}, R. and {Salvetat}, P. and {Sauvageot}, J.~L. and {Schmitt}, D. and {Sembay}, S. and {Short}, A.~D.~T. and {Spragg}, J. and {Stephen}, J. and {Str{\"u}der}, L. and {Tiengo}, A. and {Trifoglio}, M. and {Tr{\"u}mper}, J. and {Vercellone}, S. and {Vigroux}, L. and {Villa}, G. and {Ward}, M.~J. and {Whitehead}, S. and {Zonca}, E.},
        title = "{The European Photon Imaging Camera on XMM-Newton: The MOS cameras}",
      journal = {\aap},
         year = 2001,
        month = jan,
       volume = {365},
        pages = {L27-L35},
          doi = {10.1051/0004-6361:20000087},
archivePrefix = {arXiv},
       eprint = {astro-ph/0011498},
 primaryClass = {astro-ph},
       adsurl = {https://ui.adsabs.harvard.edu/abs/2001A&A...365L..27T}
}

@ARTICLE{2000ApJ...542..914W,
       author = {{Wilms}, J. and {Allen}, A. and {McCray}, R.},
        title = "{On the Absorption of X-Rays in the Interstellar Medium}",
      journal = {\apj},
         year = 2000,
        month = oct,
       volume = {542},
       number = {2},
        pages = {914-924},
          doi = {10.1086/317016},
archivePrefix = {arXiv},
       eprint = {astro-ph/0008425},
 primaryClass = {astro-ph},
       adsurl = {https://ui.adsabs.harvard.edu/abs/2000ApJ...542..914W}
}

@ARTICLE{2005AdSpR..35.1062M,
       author = {{Moldowan}, A. and {Safi-Harb}, S. and {Fuchs}, Y. and {Dubner}, G.},
        title = "{A multi-wavelength study of the western lobe of W50 powered by the galactic microquasar SS 433}",
      journal = {Advances in Space Research},
         year = 2005,
        month = jan,
       volume = {35},
       number = {6},
        pages = {1062-1065},
          doi = {10.1016/j.asr.2005.01.086},
archivePrefix = {arXiv},
       eprint = {astro-ph/0501361},
 primaryClass = {astro-ph},
       adsurl = {https://ui.adsabs.harvard.edu/abs/2005AdSpR..35.1062M}
}

@ARTICLE{2000AdSpR..25..709N,
       author = {{Namiki}, M. and {Kawai}, N. and {Kotani}, T. and {Mamauchi}, S. and {Brinkmann}, W.},
        title = "{X-Ray Lobes of W50/SS433 System}",
      journal = {Advances in Space Research},
         year = 2000,
        month = jan,
       volume = {25},
       number = {3-4},
        pages = {709-712},
          doi = {10.1016/S0273-1177(99)00827-3},
       adsurl = {https://ui.adsabs.harvard.edu/abs/2000AdSpR..25..709N}
}

@ARTICLE{2011ExA....32..193A,
       author = {{Actis}, M. and {Agnetta}, G. and {Aharonian}, F. and {Akhperjanian}, A. and {Aleksi{\'c}}, J. and {Aliu}, E. and {Allan}, D. and {Allekotte}, I. and {Antico}, F. and {Antonelli}, L.~A. and {Antoranz}, P. and {Aravantinos}, A. and {Arlen}, T. and {Arnaldi}, H. and {Artmann}, S. and {Asano}, K. and {Asorey}, H. and {B{\"a}hr}, J. and {Bais}, A. and {Baixeras}, C. and {Bajtlik}, S. and {Balis}, D. and {Bamba}, A. and {Barbier}, C. and {Barcel{\'o}}, M. and {Barnacka}, A. and {Barnstedt}, J. and {Barres de Almeida}, U. and {Barrio}, J.~A. and {Basso}, S. and {Bastieri}, D. and {Bauer}, C. and {Becerra}, J. and {Becherini}, Y. and {Bechtol}, K. and {Becker}, J. and {Beckmann}, V. and {Bednarek}, W. and {Behera}, B. and {Beilicke}, M. and {Belluso}, M. and {Benallou}, M. and {Benbow}, W. and {Berdugo}, J. and {Berger}, K. and {Bernardino}, T. and {Bernl{\"o}hr}, K. and {Biland}, A. and {Billotta}, S. and {Bird}, T. and {Birsin}, E. and {Bissaldi}, E. and {Blake}, S. and {Blanch}, O. and {Bobkov}, A.~A. and {Bogacz}, L. and {Bogdan}, M. and {Boisson}, C. and {Boix}, J. and {Bolmont}, J. and {Bonanno}, G. and {Bonardi}, A. and {Bonev}, T. and {Borkowski}, J. and {Botner}, O. and {Bottani}, A. and {Bourgeat}, M. and {Boutonnet}, C. and {Bouvier}, A. and {Brau-Nogu{\'e}}, S. and {Braun}, I. and {Bretz}, T. and {Briggs}, M.~S. and {Brun}, P. and {Brunetti}, L. and {Buckley}, J.~H. and {Bugaev}, V. and {B{\"u}hler}, R. and {Bulik}, T. and {Busetto}, G. and {Buson}, S. and {Byrum}, K. and {Cailles}, M. and {Cameron}, R. and {Canestrari}, R. and {Cantu}, S. and {Carmona}, E. and {Carosi}, A. and {Carr}, J. and {Carton}, P.~H. and {Casiraghi}, M. and {Castarede}, H. and {Catalano}, O. and {Cavazzani}, S. and {Cazaux}, S. and {Cerruti}, B. and {Cerruti}, M. and {Chadwick}, P.~M. and {Chiang}, J. and {Chikawa}, M. and {Cie{\'s}lar}, M. and {Ciesielska}, M. and {Cillis}, A. and {Clerc}, C. and {Colin}, P. and {Colom{\'e}}, J. and {Compin}, M. and {Conconi}, P. and {Connaughton}, V. and {Conrad}, J. and {Contreras}, J.~L. and {Coppi}, P. and {Corlier}, M. and {Corona}, P. and {Corpace}, O. and {Corti}, D. and {Cortina}, J. and {Costantini}, H. and {Cotter}, G. and {Courty}, B. and {Couturier}, S. and {Covino}, S. and {Croston}, J. and {Cusumano}, G. and {Daniel}, M.~K. and {Dazzi}, F. and {de Angelis}, A. and {de Cea Del Pozo}, E. and {de Gouveia Dal Pino}, E.~M. and {de Jager}, O. and {de La Calle P{\'e}rez}, I. and {de La Vega}, G. and {de Lotto}, B. and {de Naurois}, M. and {de O{\~n}a Wilhelmi}, E. and {de Souza}, V. and {Decerprit}, B. and {Deil}, C. and {Delagnes}, E. and {Deleglise}, G. and {Delgado}, C. and {Dettlaff}, T. and {di Paolo}, A. and {di Pierro}, F. and {D{\'\i}az}, C. and {Dick}, J. and {Dickinson}, H. and {Digel}, S.~W. and {Dimitrov}, D. and {Disset}, G. and {Djannati-Ata{\"\i}}, A. and {Doert}, M. and {Domainko}, W. and {Dorner}, D. and {Doro}, M. and {Dournaux}, J.-L. and {Dravins}, D. and {Drury}, L. and {Dubois}, F. and {Dubois}, R. and {Dubus}, G. and {Dufour}, C. and {Durand}, D. and {Dyks}, J. and {Dyrda}, M. and {Edy}, E. and {Egberts}, K. and {Eleftheriadis}, C. and {Elles}, S. and {Emmanoulopoulos}, D. and {Enomoto}, R. and {Ernenwein}, J.-P. and {Errando}, M. and {Etchegoyen}, A. and {Falcone}, A.~D. and {Farakos}, K. and {Farnier}, C. and {Federici}, S. and {Feinstein}, F. and {Ferenc}, D. and {Fillin-Martino}, E. and {Fink}, D. and {Finley}, C. and {Finley}, J.~P. and {Firpo}, R. and {Florin}, D. and {F{\"o}hr}, C. and {Fokitis}, E. and {Font}, Ll. and {Fontaine}, G. and {Fontana}, A. and {F{\"o}rster}, A. and {Fortson}, L. and {Fouque}, N. and {Fransson}, C. and {Fraser}, G.~W. and {Fresnillo}, L. and {Fruck}, C. and {Fujita}, Y. and {Fukazawa}, Y.},
        title = "{Design concepts for the Cherenkov Telescope Array CTA: an advanced facility for ground-based high-energy gamma-ray astronomy}",
      journal = {Experimental Astronomy},
         year = 2011,
        month = dec,
       volume = {32},
       number = {3},
        pages = {193-316},
          doi = {10.1007/s10686-011-9247-0},
archivePrefix = {arXiv},
       eprint = {1008.3703},
 primaryClass = {astro-ph.IM},
       adsurl = {https://ui.adsabs.harvard.edu/abs/2011ExA....32..193A}
}

@INPROCEEDINGS{2024icrc.confE.808Z,
       author = {{Zhang}, S. and {Wang}, Y. and {Liu}, J. and {Feng}, S. and {Yang}, M. and {Geng}, L. and {Zhang}, Y. and {LACT group}},
        title = "{Large Array of imaging atmospheric Cherenkov Telescopes (LACT): status and future plans}",
    booktitle = {38th International Cosmic Ray Conference},
         year = 2024,
        month = sep,
          eid = {808},
        pages = {808},
          doi = {10.22323/1.444.0808},
       adsurl = {https://ui.adsabs.harvard.edu/abs/2024icrc.confE.808Z}
}

@ARTICLE{2025ChPhC..49c5001Z,
       author = {{Zhang}, Zhipeng and {Yang}, Ruizhi and {Zhang}, Shoushan and {Xie}, Zhen and {Liu}, Jiali and {Yin}, Liqiao and {Wang}, Yudong and {Ma}, Lingling and {Cao}, Zhen},
        title = "{Layout optimization and performance analysis of large array of imaging atmospheric Cherenkov telescopes}",
      journal = {Chinese Physics C},
         year = 2025,
        month = mar,
       volume = {49},
       number = {3},
          eid = {035001},
        pages = {035001},
          doi = {10.1088/1674-1137/ad8e3f},
archivePrefix = {arXiv},
       eprint = {2409.14382},
 primaryClass = {astro-ph.HE},
       adsurl = {https://ui.adsabs.harvard.edu/abs/2025ChPhC..49c5001Z}
}

@ARTICLE{2022JHEAp..35...52S,
       author = {{Scuderi}, S. and {Giuliani}, A. and {Pareschi}, G. and {Tosti}, G. and {Catalano}, O. and {Amato}, E. and {Antonelli}, L.~A. and {Becerra Gonz{\`a}les}, J. and {Bellassai}, G. and {Bigongiari}, C. and {Biondo}, B. and {B{\"o}ttcher}, M. and {Bonanno}, G. and {Bonnoli}, G. and {Bruno}, P. and {Bulgarelli}, A. and {Canestrari}, R. and {Capalbi}, M. and {Caraveo}, P. and {Cardillo}, M. and {Conforti}, V. and {Contino}, G. and {Corpora}, M. and {Costa}, A. and {Cusumano}, G. and {D'A{\`\i}}, A. and {de Gouveia Dal Pino}, E. and {Della Ceca}, R. and {Escribano Rodriguez}, E. and {Falceta-Gon{\c{c}}alves}, D. and {Fermino}, C. and {Fiori}, M. and {Fioretti}, V. and {Fiorini}, M. and {Gallozzi}, S. and {Gargano}, C. and {Garozzo}, S. and {Germani}, S. and {Ghedina}, A. and {Gianotti}, F. and {Giarrusso}, S. and {Gimenes}, R. and {Giordano}, V. and {Grillo}, A. and {Grivel Gelly}, C. and {Impiombato}, D. and {Incardona}, F. and {Incorvaia}, S. and {Iovenitti}, S. and {La Barbera}, A. and {La Palombara}, N. and {La Parola}, V. and {Lamastra}, A. and {Lessio}, L. and {Leto}, G. and {Lo Gerfo}, F. and {Lodi}, M. and {Lombardi}, S. and {Longo}, F. and {Lucarelli}, F. and {Maccarone}, M.~C. and {Marano}, D. and {Martinetti}, E. and {Mereghetti}, S. and {Miccich{\'e}}, A. and {Millul}, R. and {Mineo}, T. and {Mollica}, D. and {Morlino}, G. and {Morselli}, A. and {Naletto}, G. and {Nicotra}, G. and {Pagliaro}, A. and {Parmiggiani}, N. and {Piano}, G. and {Pintore}, F. and {Poretti}, E. and {Olmi}, B. and {Rodeghiero}, G. and {Rodriguez Fernandez}, G. and {Romano}, P. and {Romeo}, G. and {Russo}, F. and {Sangiorgi}, P. and {Saturni}, F.~G. and {Schwarz}, J.~H. and {Sciacca}, E. and {Sironi}, G. and {Sottile}, G. and {Stamerra}, A. and {Tagliaferri}, G. and {Testa}, V. and {Umana}, G. and {Uslenghi}, M. and {Vercellone}, S. and {Zampieri}, L. and {Zanmar Sanchez}, R.},
        title = "{The ASTRI Mini-Array of Cherenkov telescopes at the Observatorio del Teide}",
      journal = {Journal of High Energy Astrophysics},
         year = 2022,
        month = aug,
       volume = {35},
        pages = {52-68},
          doi = {10.1016/j.jheap.2022.05.001},
archivePrefix = {arXiv},
       eprint = {2208.04571},
 primaryClass = {astro-ph.IM},
       adsurl = {https://ui.adsabs.harvard.edu/abs/2022JHEAp..35...52S}
}
\bibliographystyle{aasjournalv7}



\end{document}